\documentclass[%
 reprint,
 amsmath,amssymb,
 aps,
]{revtex4-2}

\usepackage{graphicx}% Include figure files
\usepackage{dcolumn}% Align table columns on decimal point
\usepackage{bm}% bold math
\usepackage{graphicx}
\usepackage{xcolor}
\usepackage{hyperref}
\usepackage{natbib}
\usepackage{svg}
\usepackage{amsmath}
\usepackage{amssymb}

\begin{document}

\preprint{APS/123-QED}

\title{Detecting Extragalactic Planets with Fast Radio Burst Echo Nanolensing}% Force line breaks with \\

\author{Kenneth Olibrice}
\email{kenneth8@stanford.edu}
\author{Dylan L. Jow}%
 \email{dylanjow@stanford.edu}
\affiliation{%
Kavli Institute for Particle Astrophysics \& Cosmology, Stanford University, Stanford, CA 94305, USA
}%

\date{\today}% It is always \today, today,

\begin{abstract}
Fast Radio Bursts (FRBs) are spatially and temporally compact sources at cosmological distances. As background sources, they will be uniquely sensitive to the gravitational lensing of exoplanets outside of our own galaxy. We present a method to detect extragalactic exoplanets using planetary echo nanolensing---periodic, nanosecond-scale variations in the relative time of arrival (ToA) between an initial FRB and its microlensed echo as the planet orbits its host star. For a simple binary lens model, we demonstrate that measuring the amplitude and period of these ToA fluctuations enables a determination of the exoplanet mass and orbital parameters. FRBs will be most sensitive to echo nanolensing by lenses within ten megaparsecs of the Milky Way as well as within ten megaparsecs of the FRB host, opening the door to detections of planets at a range of redshifts. We estimate that at least one in a million FRBs will exhibit this phenomenon (i.e. an optical depth of $\tau_l = 10^{-6}$). While current burst rates and low repeating-FRB fractions make a detection challenging in the immediate term, upcoming high-volume surveys and next-generation arrays (such as the SKA Phase 2) will bring the unambiguous detection of extragalactic exoplanets across cosmological redshifts into possibility.
\end{abstract}

\maketitle

%\tableofcontents

\section{\label{sec:intro}Introduction}

The evolution of planetary populations over cosmic history is poorly characterized. When did the first planets form? What was their size and distribution? When did the first rocky planets form? The answers to these questions are directly relevant to the question of when the universe could have first supported life, and how prevalent the phenomenon might be today.

To date, more than 5,000 confirmed galactic exoplanets have been discovered \citep{christiansen_nasa_2025}. These have been exclusively detected at $z=0$; or, more precisely, within our own galaxy. Only a handful of candidate extragalactic exoplanets have been reported. The first definitive detection of an exoplanet was carried out using pulsar timing \citep{WolszczanFrail1992}; however, since then, the majority of exoplanets have been discovered using radial velocity and transit methods, which are typically only sensitive to nearby planetary systems \citep{2026AJ....171..273S}. Gravitational microlensing---whereby foreground planetary systems can occasionally produce extreme magnifications of background sources---has also been used to detect a number of exoplanets \citep{Gaudi2012}. Gravitational microlensing is not distance-dependent in the same way that radial velocity measurement and transits are. Very distant planets can produce large magnifications in background sources. Likewise, the characteristic Shapiro delays induced by a strong gravitational point-lens, $\tau \sim 4 GM/c^3$, is independent of distance. 

Nevertheless, extragalactic exoplanets have yet to be conclusively detected via gravitational lensing. The background source needs to be either: 1. sufficiently spatially compact in order to be sensitive to the lensing caustics, 2. or sufficiently temporally compact in order to resolve the Shapiro delays. More precisely, in order to by highly magnified by caustics, the source must be compact relative to the Einstein angle, $\theta_E = \sqrt{4 G M /D c^2}$, where $D = D_l D_{s} /D_{ls}$ is a combination of the angular diameter distance to the lens, the source, and between the lens and source. For cosmological distances, $D \sim {\rm Gpc}$, and a Jupiter-mass planet, the Einstein angle is $10\,{\rm nas}$. The characteristic Shapiro delay for a Jupiter-mass object is $\tau \sim 10\,{\rm ns}$. 

Fast radio bursts (FRBs) are cosmological, compact, and coherent sources of radio emission. They are cosmological, meaning they are sufficiently distant to be lensed by foreground extragalactic planets \citep{2023RvMP...95c5005Z}; they are highly compact, with emission regions less than a few thousand kilometres \citep{2023RvMP...95c5005Z}; and FRBs are coherent, meaning that the timing precision that can be achieved is limited not by the burst width but by the inverse bandwidth of the observation (i.e. nanoseconds for GHz emission) \citep{2023RvMP...95c5005Z, Wucknitz2021}. Thus, FRBs are the only extragalactic sources that are sufficiently compact (spatially and temporally) to detect the gravitational lensing effect from planets in both magnification and timing. 

Candidate detections of microlensing due to extragalactic planets have already been reported: most notably, the microlensing event PA-99-N2 in M31 \citep{An2004}. In that case, the source and candidate lens were both in M31. When the lens and source are in the same galaxy, the effective distance is $D \sim D_{ls} \lesssim 10\,{\rm kpc}$, which softens the compactness requirement. However, even for the nearby M31, the individual sources (stars) are typically not resolved, and microlensing events must be identified by the temporary brightening of individual pixels which contain many unresolved sources. This makes an unambiguous determination of the properties of the lensing system challenging. \citet{Tuntsov2024} argue that lensing anomalies in the strong lensing system Q2237+0305 can be explained by a population of free floating planets in the lensing galaxy. This interpretation, however, relies on modelling of the lens matter distribution, and, in any case, cannot provide detections of individual extragalactic planets. 

In contrast, we argue that FRB lensing offers the best possibility for an unambiguous detection of extragalactic planets. One in every thousand FRBs is expected to be microlensed by a star in a foreground galaxy \citep{Connor2023}, and most stellar systems are expected to host planets \citep{2012Natur.481..167C}. As we will show, the likelihood of encountering the planetary magnification caustics is low for extragalactic lenses; thus, the primary effect is in the variation of the relative arrival times of the microlensed images. An FRB microlensed by a foreground star will be observed as an initial burst followed by a lensed echo. The relative delay between these two images will be perturbed by large planetary masses, leading to a periodic, nanosecond variation in the relative time of arrival of the echo as the planetary mass orbits its star. We refer to this phenomenon as echo nanolensing. A periodic echo nanolensing signature in a repeating, microlensed FRB may be a unique signature of extragalactic exoplanets. By measuring the amplitude and period of this variation, one can infer the mass and orbital separation of the planetary lens.

This letter is organized as follows: in Section~\ref{sec:observables}, we describe the echo nanolensing phenomenon in more detail, delineating the observables and parameters that can be inferred. In Section~\ref{sec:sensitivity}, we discuss requirements for the phenomenon to be observable. In Section~\ref{sec:rate}, we estimate the optical depth of the phenomenon, and in Section~\ref{sec:discussion}, we discuss the possibility of a detection in future FRB surveys.

\section{\label{sec:observables}Observables}

\begin{figure*}
    \centering
    \includegraphics[width=1\linewidth]{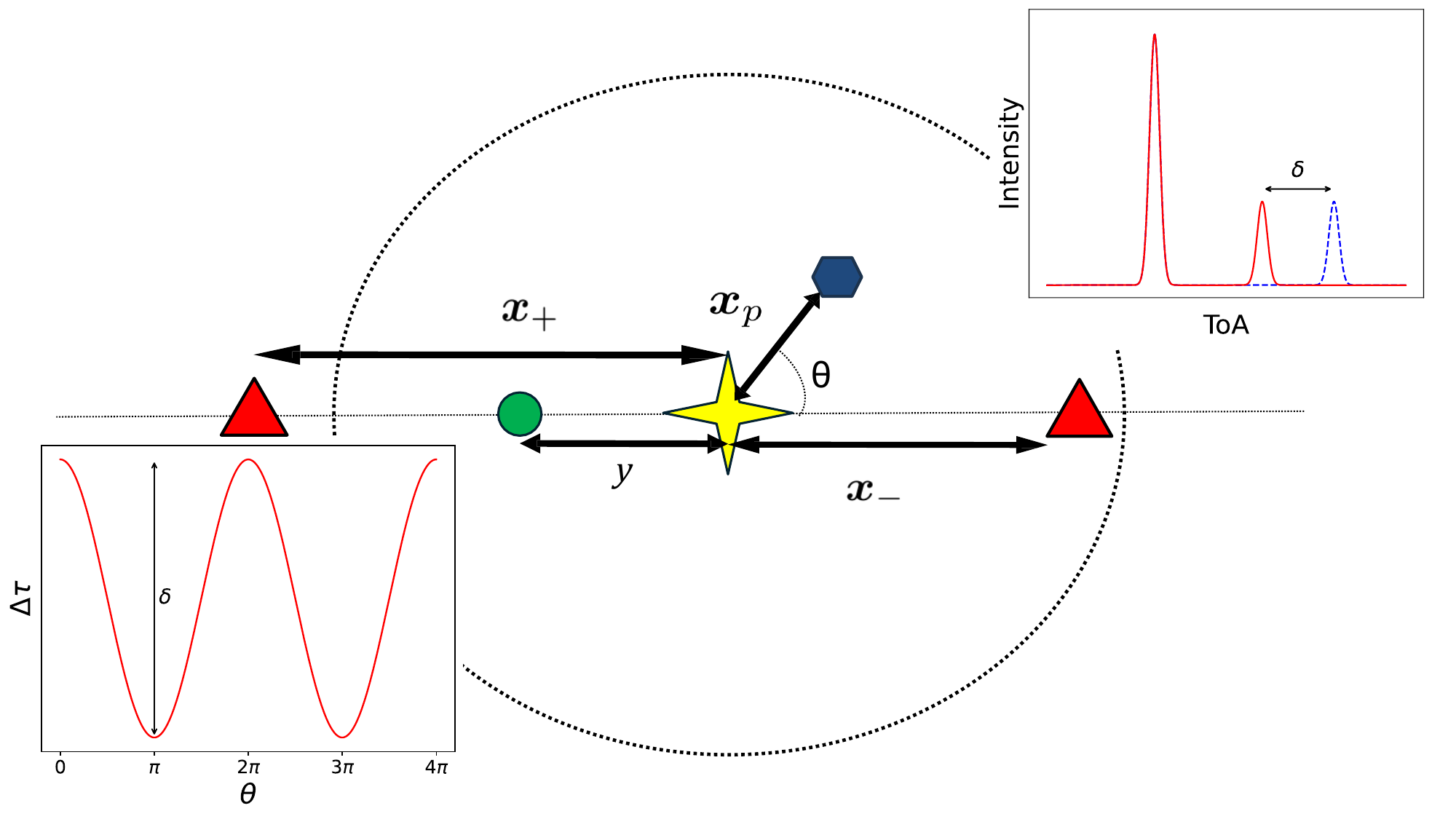}  
    \caption{A diagram of the lensing phenomenon. The background plot shows the star (yellow star), the planet (blue hexagon), and the lensed source (the green circle) at an impact parameter $y$ in the lens plane. The planet has a physical orbital separation of $a$, which, converted to an angle, and expressed in units of the Einstein angle is $d = a/D_l/\theta_E$. The two microlensed images (the red triangles) at positions $x_\pm$ are shown. The dashed circle shows the Einstein radius $\theta_E$ of the microlens. The right inset plot shows a schematic of the observed intensity as a function of time. The initial $x_+$ image arrives followed by the lensed echo (the $x_-$) image. Depending on the relative loocation of the planet in the lens plane, the ToA of the lensed echo can be perturbed by a maximum of $\delta$. The left inset plot shows the ToA of the lensed echo as a function of the orbital phase of the planet for a circular orbit. The relative ToA is periodic with the orbital phase and has an amplitude of $\delta$.}
    \label{fig:setup}
\end{figure*}

In this work, we consider a simple lensing system with two point masses: a massive stellar host and an orbiting planet. Many planets will be found in binary stellar systems and every planetary system will likely have multiple massive planets; however, we take the simple binary lens as illustrative of the echo nanolensing phenomenon that may be used to detect extragalactic exoplanets. Fig.~\ref{fig:setup} shows a diagram of the phenomenon.

The effects of the binary gravitational lens is fully contained in the Fermat potential:
\begin{align}
    T({\bm \theta}; {\bm \beta}) &= \tau_s \hat{T}({\bm x} = {\bm \theta} / \theta_E,{\bm y} = {\bm \beta} / \theta_E) \label{eq:time}, \\
     \hat{T}({\bm x}; {\bm y}) &= \frac{1}{2}|{\bm x}-{\bm y}|^2 - \ln |{\bm x}| - q\ln |{\bm x} - {\bm x}_p|,
\end{align} 

where $\theta_E = \sqrt{4 G M_s / D c^2}$ is the Einstein angle of the stellar microlens, $q = M_p/M_s$ is the ratio of the planet mass to the stellar mass, ${\bm \beta}$ is the angular position of the source relative to the optical axis, and ${\bm \theta}$ is the angular coordinate in the lens plane. The dimensionless coordinates ${\bm x} = {\bm \theta} / \theta_E$ and ${\bm y} = {\bm \beta} /\theta_E$ are normalized relative to the Einstein angle. We have also defined a dimensionless time delay, $\hat{T}$, which is related to the time delay by the characteristic timescale $\tau_s = 4 G M_s/c^3$. We choose the optical axis to be through the star so that the star is located at the origin, ${\bm x = 0}$. The position of the planet is ${\bm x}_p$, which we will take to be a function of time, ${\bm x}_p(t)$, as the planet orbits its host. 

In principle, we need to solve the full binary lens equation, ${\bm y} = \nabla T({\bm x}_i)$, for the position of the lensed images, $\left\{ {\bm x}_i\right\}$, their magnifications $\mu_i = \left| \partial {\bm y} / \partial {\bm x} ({\bm x}_i)\right|^{-1}$, and their times of arrival $\tau_i = T({\bm x}_i)$. However, since we are specifically interested in systems with a low mass ratio, $q \ll 1$, we can simplify further. Consider a system with $q = 10^{-3}$ and a separation $x_p = 0.01$ (this separation is obtained assuming a background source at infinity so that $D = D_l$. We take, as an example, a solar-mass star, $M_s = 1\,M_\odot$ at a distance $D_l \sim 1{\rm Mpc}$, with an orbital separation between the planets and star of $1\,{\rm au}$). For such a system, we can compute the areal size of the lensing caustic regions in the source plane to be $\sim 10^{-4}\,\theta^2_E$, i.e. a small fraction of the effective microlensing cross-section. Since it is only when the source is near the caustics that the effect of the planet on the magnification and image positions is large relative to the star, we will consider the effect of the planet to be perturbative, manifesting primarily as an effect on the time of arrival of the images. 

Concretely, a stellar microlens on its own produces two images, ${\bm x}_\pm$, with magnifications, $\mu_\pm$:
\begin{align}
    {\bm x}_\pm &= \pm \frac{1}{2} \frac{\bm y}{y} \left ( \sqrt{y^2 + 4} \pm y\right) \label{eq:lens}, \\
    \mu_\pm &= \frac{1}{2} \left ( \frac{y^2 + 2}{y \sqrt{y^2 + 4}} \pm 1\right ) \label{eq:mag}
\end{align}
We will take these to be the image positions and magnifications of the full binary lens system, which do not depend on the planet parameters, $q$ and ${\bm x}_p$. The image times of arrival are given by
\begin{align}
    \tau_\pm = \tau_s \hat{T}({\bm x}_\pm),
    \label{eq:taupm}
\end{align}
where we include the perturbing effect of the planet. We further justify this approach in Appendix~\ref{app:approximation}, where we perform the full binary-lens calculation and show that, for the parameters of interest, it matches our approximation. To give a sense of scale for the regime we are interested in, $\theta_E \sim 0.1\,{\rm mas}$ for a $1\,M_\odot$-lens at $D = 1\,{\rm Mpc}$. An orbital separation of $1\,{\rm au}$ at $1\,{\rm Mpc}$ subtends an angle of $\sim \mu{\rm as}$, so that $x_p \sim 10^{-2}$. As we will see, the effect we are interested in will generally require impact parameters of $y \sim 1$ for which $x_\pm \sim 1 \gg x_p$. That is, the microlensed images will typically form far from the planetary mass in the lens plane.

With the lens model written down, we can now describe the observable phenomenon. Consider a repeating FRB that is microlensed by a star with a planet orbiting it: Fig.~\ref{fig:setup} shows a schematic of the setup. Microlensing by the star produces a magnified primary image ($+$) which arrives first and a dimmer echo ($-$) for each burst of a repeating FRB. For each burst, the planet is in a different phase of its orbit, and the time delay between the initial burst and its echo varies periodically with the planetary orbit. We want to measure the amplitude of the time-varying ToAs over the entire orbit. Note that we cannot measure the absolute time delay each image experiences. Instead, we can only measure the relative times of arrival of the two images: $\Delta \tau (t) = \tau_+ - \tau_-$. This relative time delay is a function of time, varying as the planet orbits the stellar mass via ${\bm x}_p(t)$. The relative time delay is periodic as the planet orbits. We will call $\delta$ the amplitude of this periodic signal. The peak of this periodic signal occurs when the planet position ${\bm x}_p$ is closest to the lensed echo ${\bm x}_-$ in the lens plane. Likewise, the trough occurs when the planet is furthest. We take the minimum value to be an estimate of the unperturbed microlensing time delay, ${\rm min} (\Delta \tau) \approx \Delta \tau \big|_{q = 0}$. The absolute image magnifications are also not observable and we are limited to measuring the magnification ratio $\mu_- / \mu_+$. Thus, the signal is effectively characterized by four observables: the brightness ratio $\mu_- / \mu_+$, the minimum relative ToA of the echo ${\rm min}(\Delta \tau)$, the amplitude of the variation in the relative ToA $\delta$, and the period of this signal $t_o$. The first two observables do not depend on the planetary mass or separation and are given by:
\begin{align}
    &{\rm min}(\Delta \tau) = \tau_s\left ( 1+\frac{y^2}{2} - y \sqrt{y^2 + 4} + \log{\left | \frac{y-\sqrt{y^2+4}}{y+\sqrt{y^2+4}}\right |} \right )\\
    &\mu_- / \mu_+ = \frac{2 + y^2 - y\sqrt{y^2 + 4}}{2 + y(y + \sqrt{y^2 + 4})}.
    \label{eq:magratio}
\end{align}
If the distance $D$ is known, these can be directly inverted to infer the stellar mass, $M_s$, and the impact parameter, $y$.

For a face-on circular orbit, the planetary separation is given by ${\bm x}_p (t) = \frac{a}{D_l \theta_E} \left(\cos (2 \pi t/t_o), \sin(2 \pi t/t_o)\right)$, where $a$ is the radius of the orbit. The period of the time-varying ToA is the same as the orbital period, $t_o$. The maximum and minimum ToA is obtained when the planet is closest to the $+$ or $-$ image, respectively, and we can compute (see Appendix~\ref{app:cs}):
\begin{equation}
    \delta = \tau_s q \log \left| \frac{1 + d^2 + d \sqrt{y^2 + 4}}{1 + d^2 - d \sqrt{y^2 + 4}}\right|,
\end{equation}
where $d = a / D_l /\theta_E$ is the angle subtended by the orbital separation in the lens plane in units of the Einstein angle. Given that $y$ and $M_s$ can be inferred, this can be inverted to infer the orbital separation $a$ (if the distances $D_l, D_s, D_{ls}$ are known). Thus, for a face-on circular orbit, the stellar and planetary mass, the orbital period, and the orbital radius can all be determined from the observables, leveraging the Kepler relation $t_o^2 = 4 \pi^2 a^3 / GM_s$. 

In general, we expect orbits to be slightly eccentric and for not all systems to be viewed face-on. In these cases, the orbit will be projected onto an ellipse in the lens plane whose semi-major and semi-minor axes are bounded by their values in the face-on case. In the limit of small mass ratios, $\mu_-/\mu_+$ is unaffected by the change, but $\delta$ is evaluated at the orbital points nearest and farthest from the source. This means that $\delta$ is bounded by its value in the circular face-on case.

In this exposition, we have also neglected the relative motion of the source and lens, assuming a fixed value of ${\bm y}$. In reality, a stellar system at a distance $D = 1\,{\rm Mpc}$ moving with peculiar velocity $\sim 100\,{\rm km /s}$, will drift by $\sim 10\,{\rm \mu as / year}$. For a solar mass star, this translates to a change in the dimensionless impact parameter of $\Delta y \sim 0.1$ per year. This leads to a linear drift of $\sim {\rm \mu s / year}$ in the relative ToA of the two microlensed images that needs to be corrected for in order to measure the nanosecond ToA perturbations from the planetary mass.

\section{\label{sec:sensitivity}Sensitivity}

For a planetary echo nanolensing event to be detectable and for the mass to be inferred, two criteria need to be satisfied: 1. the dimmer ($-$) image (the echo) needs to be bright enough to be detected, and 2. the variation in the ToA, $\delta$, needs to be larger than the achievable timing precision. In Section~\ref{sec:rate}, we compute the optical depth for this kind of planetary echo nanolensing. The result naturally depends on the thresholds for observability that we choose. Here, we comment on our choice of thresholds.  

For the second criterion, we take the potential timing precision to be $A_{\tau} = 1\,{\rm ns}$. Thus, for an event to be observable, we require the amplitude of the ToA variation to exceed a nanosecond. Since the wavefield of FRBs can be measured directly and the raw voltage data coherently correlated between the initial burst and its lensed echo, the achievable timing precision can, in principle, be as fine as the inverse bandwidth of the observation $\Delta \nu^{-1}$. Wide bandwidth instruments such as CHORD and the DSA, with bandwidths that exceed a GHz, will be able to achieve sub-nanosecond timing precision. In practice, this is of course limited by the bandwidths of the burst themselves. Nevertheless, we take $A_{\tau} = 1\,{\rm ns}$ to be the nominal threshold. 

In order for the ToA to be relevant, the lensed echo must be detected in the first place. There is, therefore, a lower limit on the detectable magnification ratio $\mu_- / \mu_+$. For most FRB experiments, a signal-to-noise threshold of ${\rm SNR}\sim 10$ in the intensity data is typically adopted for burst detection. A lensed echo can, therefore, be de-magnified by a factor of ${\rm SNR}^{-1}$ and still be detectable (in intensity) above the noise. However, since the lensed echoes are coherent copies of the original burst, we can search for them in the raw voltage data by using the initial burst as a matched-filter template. Consider $N$ independent voltage samples $V_i = a_i +n_i$, where $a_i$ is the signal and $n_i$ is Gaussian noise with zero mean and variance $\sigma^2$. Summing the voltages across the burst gives a signal-to-noise of ${\rm SNR}_V = \rho \sqrt{N}$, where $\rho \equiv \langle a \rangle /\sigma$. Now, consider the intensity $I = V^2$. The integrated intensity signal-to-noise ratio can be calculated as ${\rm SNR}_I = \rho^2 \sqrt{N}$. Thus, we relate the voltage signal-to-noise ratio of the lensed echo to the intensity signal-to-noise ratio of the initial burst: ${\rm SNR}_V = N^{1/4} \sqrt{\mu_- {\rm SNR}_I}$. If we take ${\rm SNR_I} \sim 10$ and $N \sim 10^6$ (assuming a millisecond burst sampled at nanosecond resolution), then, requiring that ${\rm SNR}_V$ exceed unity, we find a minimum magnification of $\mu_- > A_\mu \sim 10^{-4}$. Note that many bursts will be detected with SNRs exceeding 10, resulting in a proportionately lower minimum magnification. Moreover, since our proposed measurement requires repeating FRBs, the detectability of the lensed echo could be improved with a stacking analysis over many bursts. Nevertheless, we conservatively adopt $A_\mu = 10^{-4}$ for all bursts. This is consistent with the magnification threshold used in~\citet{Leung2022}.

\section{\label{sec:rate}Optical Depth}

In this section, we calculate the probability that a given FRB will exhibit echo nanolensing by a planetary system. For a given lensing system, we define the angular cross section to be the angular area on the sky that a source can occupy to result in a detection. The requirement that the amplitude of the ToA variation exceed $A_\tau = 1\,{\rm ns}$ results in a minimum impact parameter: $y_{\rm min}$. If a source is within this minimum impact parameter, the lensed images, $x_\pm$, are too far from the origin to be perturbed by the planetary mass. For example, when $y=0$, the lensed images lie at the Einstein radius, $x_\pm = \theta_E$, which for a solar mass system at cosmological distances can be as large as thousands of ${\rm au}$---far from any orbiting planet. However, the magnification of the lensed echo decreases monotonically with the impact parameter. If the impact parameter exceeds some $y_{\rm max}$, then the lensed echo is demagnified below the magnification threshold $A_\mu = 10^{-4}$. These conditions define an annular cross section:
\begin{equation}
    \sigma = \pi \theta^2_E (y^2_{\rm max} - y^2_{\rm min}).
\end{equation}
We derive simple expressions for $y_{\rm min}$ and $y_{\rm max}$ in Appendix~\ref{app:cs}. Fig.~\ref{fig:m31_oa} shows a simple example for a Sun-Jupiter-like system in M31.

\begin{figure}
    \centering
    \includegraphics[width=\linewidth]{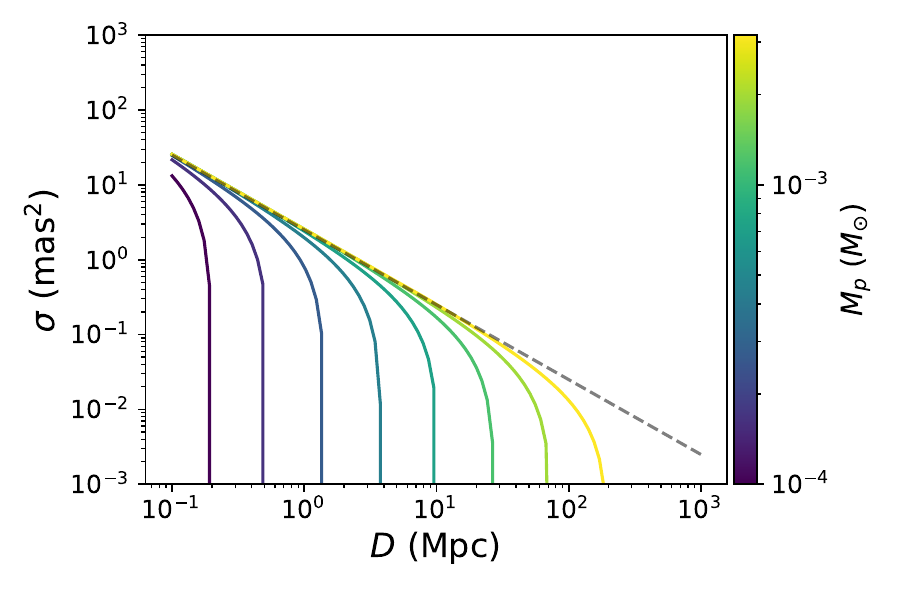}
    \caption{The echo nanolensing cross section as a function of the effective distance $D$ for $M_s = 1\,M_\odot$ and $a = 1\,{\rm au}$. The cross section is shown for different values of the planet mass, $M_p$.}
    \label{fig:csxn_distance_scaling}
\end{figure}

Fig.~\ref{fig:csxn_distance_scaling} shows the angular cross section for a system with $M_s = 1\,M_\odot$ and $a = 1\,{\rm au}$ as a function of the lens distance, $D$, for different planet masses. The cross section decreases gradually with distance, $\sigma \sim D^{-1}$, until it reaches some maximum value. The minimum impact parameter increases as a function of the effective distance, $D$, whereas $y_{\rm max}$ is only a function of $A_\mu$. Thus, there is a distance beyond which $y_{\rm min}$ exceeds $y_{\rm max}$, and the cross section vanishes. The exact distance at which this occurs depends onthe stellar mass, the planet mass, and the physical separation between the planet and its star (Eq.~\ref{eq:Dlmax}). However, as shown in Fig.~\ref{fig:csxn_distance_scaling}, this distance is below $D \lesssim 10\,{\rm Mpc}$ for the majority of the relevant parameter space for planetary lensing. Since this is a requirement on the \textit{effective} distance $D = D_l D_{ls} / D_s$, the lensing cross section is only significant for lenses either near the observer (within $\sim 10\,{\rm Mpc}$ of the Milky Way) or near the source.

\begin{figure}
    \centering
    \includegraphics[width=\linewidth]{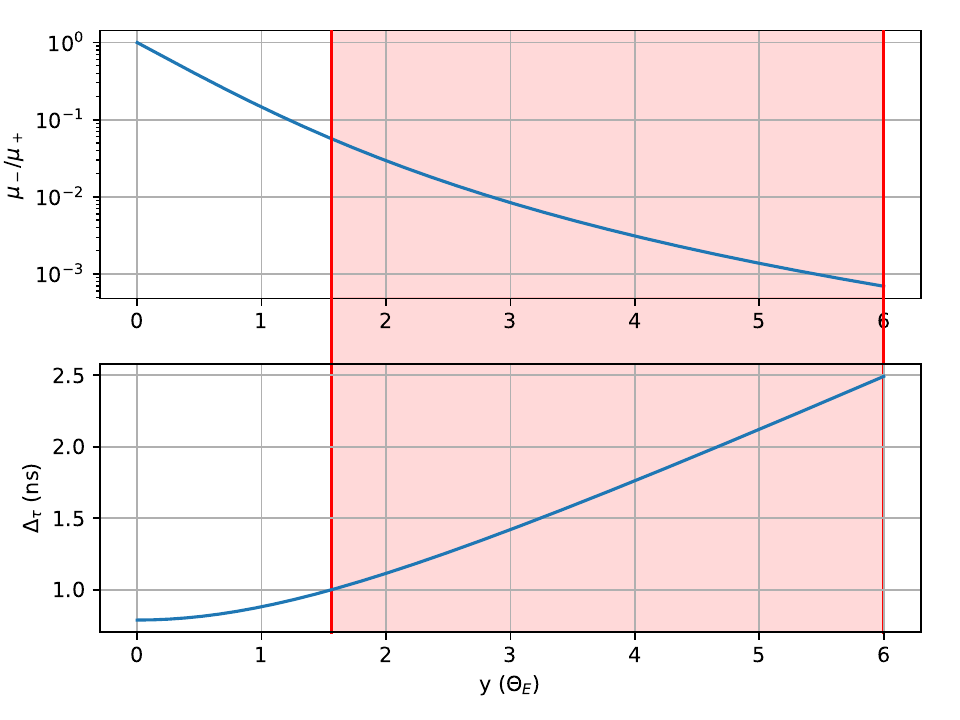}
    \caption{A visual representation of the cross section for a system with $q = 10^{-3}$, $d=10^{-2}$, and $\tau_s = 1\,{\rm \mu s}$. This corresponds roughly to a Sun-Jupiter pair located in M31. The top panel shows the magnification ratio as a function of the impact parameter, and the bottom panel shows the amplitude of the ToA variations $\delta$. The cross section is defined as the region in the lens plane for which the magnification ratio exceeds some threshold $A_\mu$ and the ToA variations exceed some threshold $A_\tau$. The red shaded region shows this region for $A_\mu = 10^{-4}$ and $\delta = 1\,{\rm ns}$.}
    \label{fig:m31_oa}
\end{figure}

\subsection{M31 Optical Depth}

Here, we will compute the optical depth contributed by the nearest major galaxy, M31. We first perform a simple order-of-magnitude estimate to get a sense of scale. Assuming a source at infinity, $D = D_l = 0.78\,{\rm Mpc}$. For a planetary system with $M_s = 1\,M_\odot$, $M_p = 10^{-3}\,M_\odot$, and $a = 1\,{\rm au}$, the angular cross section is $\sigma \sim 3\,{\rm mas}^2$. Assuming there are roughly $N \sim 10^{11}$ such systems in M31, we multiply the individual cross section by this number to obtain a total cross section, which we then divide by $4\pi\,{\rm sr}$ to obtain an optical depth of $\tau_l \sim 6 \times10^{-7}$. That is, a given FRB has a $6 \times 10^{-7}$ chance of being observably lensed by a planetary system in M31. 

A more robust estimate can be obtained by performing the integral:
\begin{equation}
    \tau_l= \frac{1}{4\pi}\int \sigma (M_s, M_p, a | D) dN_i,
\end{equation}
where 
\begin{equation}
    dN = N p(M_s, M_p, a) dM_sdM_pda.
\end{equation}
Here $p(M_s, M_p, a)$ is the probability density of a system with stellar mass, $M_s$, to have a planetary mass, $M_p$, at an orbital separation $a$. We take $N$ to be the total number of planetary systems. In the absence of a strong a priori expectation for this distribution, we adopt a simple prescription informed by the NASA Exoplanet Archive \cite{christiansen_nasa_2025}, which catalogs all confirmed exoplanets. For simplicity, we assume independence so that $p(M_s, M_p, a) = p(M_s) p(M_p) p(a)$ and fit a simple log-normal distribution to the empirical distributions from the archive. The result is that we take $\log_{10}(M_s/M_\odot) \sim \mathcal{N}(-0.04, 0.16)$ and $\log_{10}(a/{\rm au}) \sim \mathcal{N}(-0.88,0.62)$. The empirical distribution of planetary masses is bimodal, so we fit a two component Gaussian mixture model, with peaks at $\log_{10} (M_p /M_\odot) = -4.77$ and $-3.19$; widths of $0.16$ and $0.34$; and relative weights of $0.66$ and $0.34$, respectively. The overall normalization is fixed such that $M_{\rm M31} = N \int p(M_s) M_s dM_s$, where $M_{\rm M31}$ is the mass of the stellar disk. This yields $N \sim 10^{11}$. Performing the integral, we obtain $\tau_l = 10^{-7}$ for M31, roughly in accordance with our initial estimate. 

We arrived at this result assuming that the distributions of stellar mass, planet mass, and orbital separation are uncorrelated. This is unlikely to be the case, and, indeed, these parameters exhibit correlations in NASA Exoplanet Archive from which we drew the empirical distributions. However, these correlations are likely driven by selection effects peculiar to the different methods of exoplanet detection, rather than reflective of intrinsic correlations. Nevertheless, to test the sensitivity of our result to our assumption, we perform the integral adopting a Kernel density estimation for the empirical distributions of the three parameters. We find the result is similar, but slightly larger. We expect that realistic intrinsic correlations will also lead to increases in the estimated optical depth, driven by the fact that larger massive planets form preferentially beyond the snow line at larger orbital separations \citep{2009A&A...501.1139M}, leading to larger values of $\delta$. 

Note that in the preceding, we have, for the sake of simplicity and clarity of exposition, assumed a binary lens model, with one star and one planet. However, stellar binaries are ubiquitous \citep{Kareem2024NewAR..9801694E}. The effect of a binary companion is to produce three or five microlensed images. This may actually enhance the sensitivity to planetary echo nanolensing as only one of the images needs to be close enough to the planet to feel its perturbing effect.

\subsection{All-sky optical depth}

In the previous section, we computed the optical depth for a single nearby galaxy, M31. As we have argued, galaxies within $D < 10\,{\rm Mpc}$ will contribute to the optical depth. To estimate the total optical depth for galaxies within $10\,{\rm Mpc}$ of the Milky Way, we note that the cross section scales as $\sigma \sim D^{-1}$ (as shown in Fig.~\ref{fig:csxn_distance_scaling}). We fix $\sigma(D=0.78\,{\rm Mpc})=4\pi \times10^{-7}$ and assume that there are $N_{\rm gal} \sim 10^2$ M31-like galaxies within $10\,{\rm Mpc}$, distributed uniformly in space so that $p(D) \propto D^2$. Then, integrating $\int \sigma(D) p(D) dD$ out to $10\,{\rm Mpc}$, we arrive at $\tau_l = 10^{-6}$. Note that because the number of contributing galaxies increases faster with distance than the cross section decreases (up to some maximum distance), the optical depth is dominated by more distance galaxies, up to some maximum limit. 

% Since the cross section grows as the distance decreases, dwarf galaxies orbiting the Milky Way may also contribute significantly to the optical depth. We estimate the optical depth contributed by the Large Magellanic Cloud by adopting the same distribution $p(M_s, M_p, a)$ as before, but normalizing the total mass to $10^9\,M_\odot$. The result is an optical depth $\tau_l = 4\cdot10^{-8}$ for the LMC. While there are tens of dwarf galaxies within $\sim 100\,{\rm kpc}$ of the Milky Way, they are all at least an order of magnitude less massive than the LMC. Thus, the largest contributor to the optical depth will be the massive galaxies within $D < 10\,{\rm Mpc}$ of the Milky Way.

While we have focused here on nearby lenses such that $D \approx D_l < 10\,{\rm Mpc}$, the reciprocity of ray optics means that we could equally have considered lenses close to the source, such that $D \approx D_{ls} < 10\,{\rm Mpc}$. Assuming the local environments (within 10\,{\rm Mpc}) of FRB sources are roughly analogous to the Milky Way, then the galaxies local to the FRB source will contribute a similar magnitude to the optical depth. This opens up an avenue for detecting planets at non-zero redshifts, potentially as far as $z>2$.

At these redshifts, cosmological time delays become important and the form of \eqref{eq:time} gains a non-negligible prefactor of $1+z_L$, where $z_L$ is the redshift of the lens.

We note that the estimated optical depth reported here is properly interpreted as a lower bound. The lensing cross section scales with the magnification threshold, $\sigma \sim A_\mu^{-1/2}$, for which we have chosen $A_\mu = 10^{-4}$, which is the minimum requirement for all bursts detected above a signal-to-noise ratio of $10$ in intensity to have a detectable microlensed echo. However, since ${\rm SNR} \sim 10$ is typically the \textit{minimum} threshold for an FRB to be considered detected, almost all FRBs will be detected above this threshold.

\section{\label{sec:discussion}Discussion}

While we have focused on a particular effect in this paper, for completeness, we will list the three distinct ways extragalactic exoplanets may imprint a lensing signature on FRBs:
\begin{enumerate}
    \item Periodic nanosecond perturbations in the ToA of a microlensed echo.
    \item Large magnifications due to caustic crossings of planetary systems in the host galaxy.
    \item Diffractive lensing from free-floating planets.
\end{enumerate}
The first is the focus of this paper, and we have estimated that at least one in a million FRBs will exhibit detectable effects. The second effect is irrelevant for lenses outside of the Milky Way or host galaxy, as we have shown that the angular area of the caustic regions are negligible for $D \gtrsim {\rm Mpc}$. However, for planetary systems within the FRB host galaxy for which $D \approx D_{ls} \lesssim 10\,{\rm kpc}$, the lensing geometry is analogous to planetary microlensing within the Milky Way. This is a well-studied phenomenon in the context of our own galaxy, and, hence, is not the focus of this paper. \citet{Mroz2019ApJS..244...29M} estimate the microlensing optical depth towards the galactic centre to be $\tau_l \sim 10^{-6}$---roughly the same order as the estimated optical depth for the nanosecond ToA perturbations. However, most FRBs will not be viewed through the centre of their own host galaxies. The third effect refers to the fact that the Schwarzschild radii of planets are on the same order as the centimetre-wavelengths of FRBs. Thus, isolated, free-floating planets will imprint a diffractive interference pattern on FRBs, as opposed to forming distinct images. The prevalence of this effect is hard to estimate as it relies on the highly uncertain population of FFPs, but may also have a similar optical depth: $\tau_l \sim 10^{-6}$ \citep{Jow2020MNRAS.497.4956J}.

Thus, we argue that at least one in a million FRBs will exhibit lensing effects from extragalactic exoplanets. Future FRB experiments such as the DSA, CHORD, and BURSTT will enable us to detect more than 10,000 FRBs a year \citep{DSA2019BAAS...51g.255H, CHORD2019clrp.2020...28V, BURSTT2022PASP..134i4106L}. However, since detecting the periodic signal of planetary echo nanolensing will require a repeating FRB, and since only a few percent of FRBs are known to regularly repeat, it will be unlikely that extragalactic exoplanet lensing will be detected with current or near-term capabilities. Nevertheless, large field-of-view phased arrays such as BURSTT, and proposed experiments such as CASPA and CASM \citep{Luo2024PASA...41..109L, Connor2023} are highly scalable and may potentially reach hundreds of FRBs per day. The SKA Phase 2 is projected to detect thousands of FRBs per day, or millions per year, bringing a detection into the realm of possibility \citep{SKArate2020MNRAS.497.4107H}. Moreover, while we have estimated an all-sky lensing optical depth, the probability of detection can be significantly improved with a careful search strategy. FRBs that are discovered to be microlensed by stars in nearby galaxies may be followed up with higher sensitivity instruments, such as FAST.  

We also emphasize that while these events may remain rare for the foreseeable future, FRB lensing is likely the only method of reliably identifying extragalactic exoplanets due to the lack of other compact cosmological sources. Moreover, since the lensing phenomenon is as sensitive to lenses within tens of megaparsecs of the Milky Way as to lenses within tens of megaparsecs of the source, FRB echo nanolensing may be used to detect planets at non-zero redshifts.

\section{\label{sec:conclusion}Conclusion}

FRBs are compact, cosmological sources which may be timed to nanosecond precision. Thus, they may be sensitive, as background sources, to gravitational lensing by extragalactic exoplanets. In this work, we have illustrated the gravitational echo nanolensing effect---a periodic nanosecond perturbation on the arrival times of a microlensed echo---due to a planetary mass orbiting a microlenisng star. We have argued that FRBs are uniquely sensitive to this effect, which provides the only pathway to an unambiguous detection of an extragalactic exoplanet. Moreover, this approach is sensitive to both planets at zero redshift (within ten megaparsecs of the Milky Way) and planets at non-trivial redshifts (within ten megaparsecs of the FRB source). As new experiments bring us into an era of millions of FRBs, detecting extragalactic exoplanets across a range of redshifts may become a possibility.

\begin{acknowledgments}
We thank Artem Tuntsov, Olaf Wucknitz, and Roger Blandford for helpful discussions and comments. D.L.J. is a Kavli Fellow supported by the Kavli Foundation. 
\end{acknowledgments}

\appendix

\section{\label{app:approximation}Binary-lens approximation}

In this section, we will justify our simplified perturbative model of planetary echo nanolensing. In principle, a binary lens (a star and planetary companion) will generate three or five images, depending on the lens configuration (\cite{Gaudi2012}). However, we are explicitly considering systems for which $q \ll 1$ and $x_p \ll 1$. The caustic regions for such systems are vanishingly small, and, generically only three images are produced. The brightest two images can be identified with the single microlensed images, $x_\pm$, whereas the third is typically de-magnified by many orders of magnitude. 

\begin{figure}
    \centering
    \includegraphics[width=\linewidth]{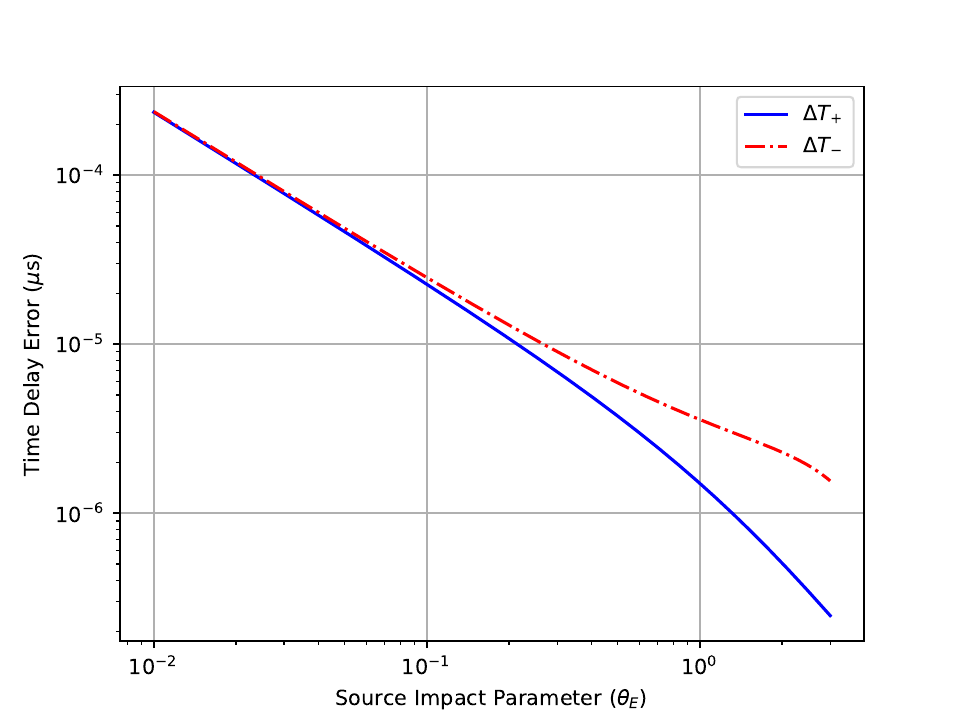}
    \caption{The error introduced by treating a binary lens as a single lens represented as the absolute difference plotted for the case $q=10^{-2}$, $x_p=10^{-3}$ representing a super-Jupiter-Sun pair with a typical projected separation at 1 Mpc.}
    \label{fig:tderror}
\end{figure}

\begin{figure}
    \centering
    \includegraphics[width=\linewidth]{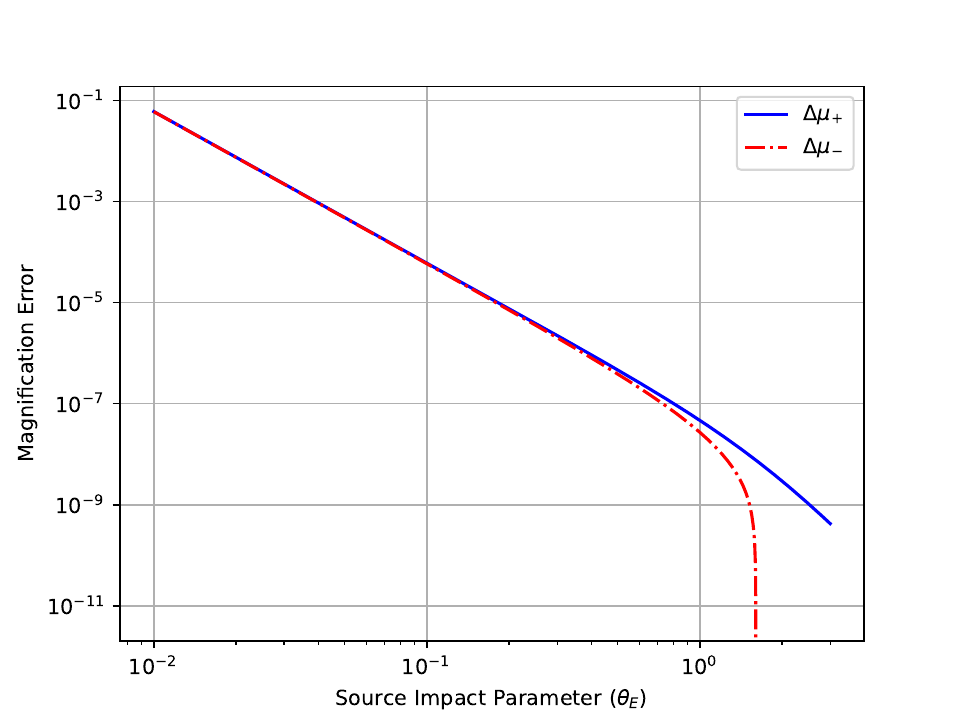}
    \caption{The error in magnification for the same system as \ref{fig:tderror}.}
    \label{fig:magerror}
\end{figure}

Consider a system with $M_s = 1\,M_\odot$, $D = 1\,{\rm Mpc}$, $q = 10^{-2}$, and $x_p = 10^{-3}$. Fig.~\ref{fig:tderror} shows the relative error between the ToAs of the two microlens images, $x_\pm$, and the two brightest images computed in the full binary lens case, as a function of the source impact parameter. As can be seen, the relative error is small, except as the source approaches the caustic region $y \to 0$. In particular, the relative error is less than a nanosecond for all $y > 10^{-2}$. Thus, the error introduced by our approximation is less than the nanosecond effect we are looking for except for very small impact parameters. Fig.~\ref{fig:magerror} shows the relative error on the magnifications. Outside of a small region near the origin, both the ToA and magnification errors are smaller than their associated thresholds ($A_\tau = 1\,{\rm ns}$ and $A_\mu =10^{-4}$, respectively). As $y \to 0$,the errors increase without bound due to the presence of the central caustic in the single lens geometry that is "smeared out" in the binary lens geometry (see \cite{Gaudi2012} for an extended discussion on caustics). However, the cross section for this is much smaller than the cross sections for the nanolensing effect.

We must also justify the use of Eq.~\ref{eq:taupm}, where we assume that the perturbing effect of the planet on the relative ToA is dominated by the potential term $q \log |{\bm x}_\pm - {\bm x}_p|$ in $\hat{T}$ rather than the additional geometric delay due to the fact that the actual image positions are slightly perturbed by the presence of the planet. That is, consider ${\bm x} = {\bm x}_\pm + \delta {\bm x}$. We want to show that $\hat{T}({\bm x}_\pm + \delta {\bm x}) = \hat{T}({\bm x}_\pm) + \mathcal{O}(q^2)$. For simplicity, we will consider the case where ${\bm y}$ lies along the same axis as the separation vector between the planet and star so that all of the expressions simplify to simple one-dimensional expressions. In this case, the full lens equation is
\begin{equation}
    y = x - \frac{1}{x} - \frac{q}{x - x_p}
    \label{eq:fullens}
\end{equation}
Letting $x = x_\pm + \delta x$ and noting $x_\pm$ solve the unperturbed lens equation, $y = x_\pm - \frac{1}{x_\pm}$, we can expand Eq.~\ref{eq:fullens} to obtain (neglecting $\mathcal{O}(\delta x^2)$ terms):
\begin{equation}
    \delta x \approx \frac{q}{(1 + x_\pm^{-2})(x_\pm - x_p)}.
\end{equation}
Expanding the time delay, we obtain
\begin{align}
\begin{split}
    &\hat{T}(x_\pm + \delta x)  \\
    &= \hat{T}(x_\pm) - \delta x \left( \frac{q}{x_\pm -x_p} \right) \\
    &+ \frac{\delta x^2}{2} \left( 1 + x_\pm^{-2} + \frac{q}{(x_\pm-x_p)^2}\right) + \mathcal{O}(\delta x^3), \\
    &= \hat{T}(x_\pm) + \mathcal{O}(q^2).
\end{split}
\end{align}

% \DLJ{The description here should be more thorough. Specifically, you should describe all of the images in the full binary lens configuration (i.e. there are three). One of them is very dim. The other two match the two in the single lens picture (which we name + and -). We want to know if the magnifications and time delays of the single-lens images is a good approximation for the full binary lens with the parameters of interest to us. Also, you want to interpret the plots for the reader. Namely, the absolute error does not exceed more than a nanosecond except as one approaches the central caustic region. But since the central caustic region is so small, it doesn't matter. Maybe refer to the next Appendix to foreshadow the fact that $y_{\rm min} \gtrsim 1$ anyway, and so errors at small $y$ are irrelevant for us.} 

\section{\label{app:cs}Cross Section Calculation}

The cross section is defined as the angular area around a lensing system wherein lensing events will be detectable. In our case, detectability is dictated entirely by the magnification ratio of the two images and the relative ToA amplitude (see \ref{sec:observables}). In this section we will derive approximate expressions for both and furthermore quantify the criteria for observability.

The first condition for observability is that the amplitude of the periodic variation in the arrival time of the two microlensed images must exceed some threshold $A_\tau$. The largest difference in arrival times occur when the planet, in the lens plane, is either closest to the source or furthest from the source. Consider, without loss of generality, a source at ${\bm y} = y \hat{x}$. The images form at ${\bm x}_\pm = x_\pm \hat{x}$. Now consider a planet with a circular orbit and orbital separation $d = a/D_l/\theta_E$: the planet is closest to the source when ${\bm x}_p = d \hat{x}$ and furthest when ${\bm x}_p = -d \hat{x}$. Thus, we can compute the amplitude of the variation in the relative ToAs as
\begin{align}
\begin{split}  
\delta = &\tau_s \Big[ \left(\hat{T}(x_+; y) - \hat{T}(x_-;y)\right)_{{\bm x}_p = d\hat{x}} \\
&- \left(\hat{T}(x_+; y) - \hat{T}(x_-;y)\right)_{{\bm x}_p = -d\hat{x}}) \Big].
\end{split}
\end{align}
This evaluates to 
\begin{equation}
    \delta / \tau_s = q \log \left |\frac{1+d^2+d\sqrt{y^2+4}}{1+d^2-d\sqrt{y^2+4}} \right |.
\end{equation}
When $d \ll 1$ and $y \ll \sqrt{d^{-2}+4}$, we can expand to first order:
\begin{equation}
    \delta / \tau_s \approx 2 q d \sqrt{y^2 + 4}.
\end{equation}

The second condition for observability is that the magnification ratio of the images must exceed some threshold. The magnification ratio can be computed using Eq.~\ref{eq:magratio}. For large values of $y$, this can be expanded to leading order as $\mu_- / \mu_+ = 1/y^4$.

Note that $\delta$ is an increasing function of $y$ whereas $\mu_-$ is a decreasing function of $y$. Therefore, in order for the planetary echo nanolensing effect to be observable, the source must be at an impact parameter that exceeds a minimum $y_\mathrm{min}$ (such that $\delta> A_\tau$) and is less than a maximum $y_\mathrm{max}$ (such that $\mu_-/\mu_+ > A_\mu$). This defines an annular cross-sectional area:
\begin{equation}
    \sigma = \pi \theta_E^2(y_\mathrm{max}^2-y_\mathrm{min}^2),
\end{equation}
where
\begin{align}
    y_{\mathrm{min}} &\approx \sqrt{\left ( \frac{A_\tau}{2 \tau_p d}\right )^2 - 4} \\
    y_\mathrm{max} &\approx A_\mu^{-\frac{1}{4}}.
\end{align}
Here we have defined the characteristic Shaprio delay of the planetary mass:
\begin{equation}
    \tau_p = \frac{4 G M_p}{c^3}.
\end{equation}

The forms of $y_\mathrm{min},y_\mathrm{max}$ given above also give an upper bound on the distance from Earth for detectable exoplanets. Namely, $y_{\rm max}$ remains fixed, while $y_{\rm min}$ increases without bound with $D$. Thus, there is a $D^{\rm max}$ at which $y_{\rm min} = y_{\rm max}$ and the cross section vanishes. This maximum distance can be computed: 
\begin{equation}
    D^{\mathrm{max}} = \frac{4 \tau_p^2 a^2}{A^2_\tau \tau_s c} (A^{-1/2}_\mu + 4)
    \label{eq:Dlmax}
\end{equation}
For a mass ratio of $10^{-3}$ orbiting at 1 AU and the thresholds adopted in Section~\ref{sec:sensitivity}, this yields a distance of roughly $20$ Mpc. In other words, planetary echo nanolensing is only observable for lenses within $20\,{\rm Mpc}$ of the Milky Way (for a source at infinity) or within $20\,{\rm Mpc}$ os the source. 

\bibliography{apssamp}% Produces the bibliography via BibTeX.

\providecommand{\noopsort}[1]{}\providecommand{\singleletter}[1]{#1}%
\begin{thebibliography}{20}%
\makeatletter
\providecommand \@ifxundefined [1]{%
 \@ifx{#1\undefined}
}%
\providecommand \@ifnum [1]{%
 \ifnum #1\expandafter \@firstoftwo
 \else \expandafter \@secondoftwo
 \fi
}%
\providecommand \@ifx [1]{%
 \ifx #1\expandafter \@firstoftwo
 \else \expandafter \@secondoftwo
 \fi
}%
\providecommand \natexlab [1]{#1}%
\providecommand \enquote  [1]{``#1''}%
\providecommand \bibnamefont  [1]{#1}%
\providecommand \bibfnamefont [1]{#1}%
\providecommand \citenamefont [1]{#1}%
\providecommand \href@noop [0]{\@secondoftwo}%
\providecommand \href [0]{\begingroup \@sanitize@url \@href}%
\providecommand \@href[1]{\@@startlink{#1}\@@href}%
\providecommand \@@href[1]{\endgroup#1\@@endlink}%
\providecommand \@sanitize@url [0]{\catcode `\\12\catcode `\$12\catcode `\&12\catcode `\#12\catcode `\^12\catcode `\_12\catcode `\%12\relax}%
\providecommand \@@startlink[1]{}%
\providecommand \@@endlink[0]{}%
\providecommand \url  [0]{\begingroup\@sanitize@url \@url }%
\providecommand \@url [1]{\endgroup\@href {#1}{\urlprefix }}%
\providecommand \urlprefix  [0]{URL }%
\providecommand \Eprint [0]{\href }%
\providecommand \doibase [0]{https://doi.org/}%
\providecommand \selectlanguage [0]{\@gobble}%
\providecommand \bibinfo  [0]{\@secondoftwo}%
\providecommand \bibfield  [0]{\@secondoftwo}%
\providecommand \translation [1]{[#1]}%
\providecommand \BibitemOpen [0]{}%
\providecommand \bibitemStop [0]{}%
\providecommand \bibitemNoStop [0]{.\EOS\space}%
\providecommand \EOS [0]{\spacefactor3000\relax}%
\providecommand \BibitemShut  [1]{\csname bibitem#1\endcsname}%
\let\auto@bib@innerbib\@empty
%</preamble>
\bibitem [{\citenamefont {Christiansen}\ \emph {et~al.}(2025)\citenamefont {Christiansen}, \citenamefont {McElroy}, \citenamefont {Harbut}, \citenamefont {Ciardi}, \citenamefont {Crane}, \citenamefont {Good}, \citenamefont {Hardegree-Ullman}, \citenamefont {Kesseli}, \citenamefont {Lund}, \citenamefont {Lynn}, \citenamefont {Muthiar}, \citenamefont {Nilsson}, \citenamefont {Oluyide}, \citenamefont {Papin}, \citenamefont {Rivera}, \citenamefont {Swain}, \citenamefont {Susemiehl}, \citenamefont {Tam}, \citenamefont {Eyken},\ and\ \citenamefont {Beichman}}]{christiansen_nasa_2025}%
  \BibitemOpen
  \bibfield  {author} {\bibinfo {author} {\bibfnamefont {J.~L.}\ \bibnamefont {Christiansen}}, \bibinfo {author} {\bibfnamefont {D.~L.}\ \bibnamefont {McElroy}}, \bibinfo {author} {\bibfnamefont {M.}~\bibnamefont {Harbut}}, \bibinfo {author} {\bibfnamefont {D.~R.}\ \bibnamefont {Ciardi}}, \bibinfo {author} {\bibfnamefont {M.}~\bibnamefont {Crane}}, \bibinfo {author} {\bibfnamefont {J.}~\bibnamefont {Good}}, \bibinfo {author} {\bibfnamefont {K.~K.}\ \bibnamefont {Hardegree-Ullman}}, \bibinfo {author} {\bibfnamefont {A.~Y.}\ \bibnamefont {Kesseli}}, \bibinfo {author} {\bibfnamefont {M.~B.}\ \bibnamefont {Lund}}, \bibinfo {author} {\bibfnamefont {M.}~\bibnamefont {Lynn}}, \bibinfo {author} {\bibfnamefont {A.}~\bibnamefont {Muthiar}}, \bibinfo {author} {\bibfnamefont {R.}~\bibnamefont {Nilsson}}, \bibinfo {author} {\bibfnamefont {T.}~\bibnamefont {Oluyide}}, \bibinfo {author} {\bibfnamefont {M.}~\bibnamefont {Papin}}, \bibinfo {author} {\bibfnamefont {A.}~\bibnamefont {Rivera}}, \bibinfo {author} {\bibfnamefont
  {M.}~\bibnamefont {Swain}}, \bibinfo {author} {\bibfnamefont {N.~D.}\ \bibnamefont {Susemiehl}}, \bibinfo {author} {\bibfnamefont {R.}~\bibnamefont {Tam}}, \bibinfo {author} {\bibfnamefont {J.~v.}\ \bibnamefont {Eyken}},\ and\ \bibinfo {author} {\bibfnamefont {C.}~\bibnamefont {Beichman}},\ }\href {https://doi.org/10.48550/arXiv.2506.03299} {\bibinfo {title} {The {NASA} {Exoplanet} {Archive} and {Exoplanet} {Follow}-up {Observing} {Program}: {Data}, {Tools}, and {Usage}}} (\bibinfo {year} {2025}),\ \bibinfo {note} {arXiv:2506.03299 [astro-ph.EP]}\BibitemShut {NoStop}%
\bibitem [{\citenamefont {{Wolszczan}}\ and\ \citenamefont {{Frail}}(1992)}]{WolszczanFrail1992}%
  \BibitemOpen
  \bibfield  {author} {\bibinfo {author} {\bibfnamefont {A.}~\bibnamefont {{Wolszczan}}}\ and\ \bibinfo {author} {\bibfnamefont {D.~A.}\ \bibnamefont {{Frail}}},\ }\bibfield  {title} {\bibinfo {title} {{A planetary system around the millisecond pulsar PSR1257 + 12}},\ }\href {https://doi.org/10.1038/355145a0} {\bibfield  {journal} {\bibinfo  {journal} {NAT}\ }\textbf {\bibinfo {volume} {355}},\ \bibinfo {pages} {145} (\bibinfo {year} {1992})}\BibitemShut {NoStop}%
\bibitem [{\citenamefont {{Schap}}\ \emph {et~al.}(2026)\citenamefont {{Schap}}, \citenamefont {{Dittmann}},\ and\ \citenamefont {{Lada}}}]{2026AJ....171..273S}%
  \BibitemOpen
  \bibfield  {author} {\bibinfo {author} {\bibfnamefont {W.}~\bibnamefont {{Schap}}}, \bibinfo {author} {\bibfnamefont {J.}~\bibnamefont {{Dittmann}}},\ and\ \bibinfo {author} {\bibfnamefont {E.}~\bibnamefont {{Lada}}},\ }\bibfield  {title} {\bibinfo {title} {{Searching for Extragalactic Exoplanets: A Survey of the Sagittarius Dwarf Galaxy Stream with TESS}},\ }\href {https://doi.org/10.3847/1538-3881/ae4345} {\bibfield  {journal} {\bibinfo  {journal} {AJ}\ }\textbf {\bibinfo {volume} {171}},\ \bibinfo {eid} {273} (\bibinfo {year} {2026})},\ \Eprint {https://arxiv.org/abs/2602.15105} {arXiv:2602.15105 [astro-ph.EP]} \BibitemShut {NoStop}%
\bibitem [{\citenamefont {{Gaudi}}(2012)}]{Gaudi2012}%
  \BibitemOpen
  \bibfield  {author} {\bibinfo {author} {\bibfnamefont {B.~S.}\ \bibnamefont {{Gaudi}}},\ }\bibfield  {title} {\bibinfo {title} {{Microlensing Surveys for Exoplanets}},\ }\href {https://doi.org/10.1146/annurev-astro-081811-125518} {\bibfield  {journal} {\bibinfo  {journal} {ARAA}\ }\textbf {\bibinfo {volume} {50}},\ \bibinfo {pages} {411} (\bibinfo {year} {2012})}\BibitemShut {NoStop}%
\bibitem [{\citenamefont {{Zhang}}(2023)}]{2023RvMP...95c5005Z}%
  \BibitemOpen
  \bibfield  {author} {\bibinfo {author} {\bibfnamefont {B.}~\bibnamefont {{Zhang}}},\ }\bibfield  {title} {\bibinfo {title} {{The physics of fast radio bursts}},\ }\href {https://doi.org/10.1103/RevModPhys.95.035005} {\bibfield  {journal} {\bibinfo  {journal} {Reviews of Modern Physics}\ }\textbf {\bibinfo {volume} {95}},\ \bibinfo {eid} {035005} (\bibinfo {year} {2023})},\ \Eprint {https://arxiv.org/abs/2212.03972} {arXiv:2212.03972 [astro-ph.HE]} \BibitemShut {NoStop}%
\bibitem [{\citenamefont {{Wucknitz}}\ \emph {et~al.}(2021)\citenamefont {{Wucknitz}}, \citenamefont {{Spitler}},\ and\ \citenamefont {{Pen}}}]{Wucknitz2021}%
  \BibitemOpen
  \bibfield  {author} {\bibinfo {author} {\bibfnamefont {O.}~\bibnamefont {{Wucknitz}}}, \bibinfo {author} {\bibfnamefont {L.~G.}\ \bibnamefont {{Spitler}}},\ and\ \bibinfo {author} {\bibfnamefont {U.-L.}\ \bibnamefont {{Pen}}},\ }\bibfield  {title} {\bibinfo {title} {{Cosmology with gravitationally lensed repeating fast radio bursts}},\ }\href {https://doi.org/10.1051/0004-6361/202038248} {\bibfield  {journal} {\bibinfo  {journal} {AAP}\ }\textbf {\bibinfo {volume} {645}},\ \bibinfo {eid} {A44} (\bibinfo {year} {2021})},\ \Eprint {https://arxiv.org/abs/2004.11643} {arXiv:2004.11643 [astro-ph.CO]} \BibitemShut {NoStop}%
\bibitem [{\citenamefont {{An}}\ \emph {et~al.}(2004)\citenamefont {{An}}, \citenamefont {{Evans}}, \citenamefont {{Kerins}}, \citenamefont {{Baillon}}, \citenamefont {{Calchi Novati}}, \citenamefont {{Carr}}, \citenamefont {{Cr{\'e}z{\'e}}}, \citenamefont {{Giraud-H{\'e}raud}}, \citenamefont {{Gould}}, \citenamefont {{Hewett}}, \citenamefont {{Jetzer}}, \citenamefont {{Kaplan}}, \citenamefont {{Paulin-Henriksson}}, \citenamefont {{Smartt}}, \citenamefont {{Tsapras}}, \citenamefont {{Valls-Gabaud}},\ and\ \citenamefont {{Point-Agape Collaboration}}}]{An2004}%
  \BibitemOpen
  \bibfield  {author} {\bibinfo {author} {\bibfnamefont {J.~H.}\ \bibnamefont {{An}}}, \bibinfo {author} {\bibfnamefont {N.~W.}\ \bibnamefont {{Evans}}}, \bibinfo {author} {\bibfnamefont {E.}~\bibnamefont {{Kerins}}}, \bibinfo {author} {\bibfnamefont {P.}~\bibnamefont {{Baillon}}}, \bibinfo {author} {\bibfnamefont {S.}~\bibnamefont {{Calchi Novati}}}, \bibinfo {author} {\bibfnamefont {B.~J.}\ \bibnamefont {{Carr}}}, \bibinfo {author} {\bibfnamefont {M.}~\bibnamefont {{Cr{\'e}z{\'e}}}}, \bibinfo {author} {\bibfnamefont {Y.}~\bibnamefont {{Giraud-H{\'e}raud}}}, \bibinfo {author} {\bibfnamefont {A.}~\bibnamefont {{Gould}}}, \bibinfo {author} {\bibfnamefont {P.}~\bibnamefont {{Hewett}}}, \bibinfo {author} {\bibfnamefont {P.}~\bibnamefont {{Jetzer}}}, \bibinfo {author} {\bibfnamefont {J.}~\bibnamefont {{Kaplan}}}, \bibinfo {author} {\bibfnamefont {S.}~\bibnamefont {{Paulin-Henriksson}}}, \bibinfo {author} {\bibfnamefont {S.~J.}\ \bibnamefont {{Smartt}}}, \bibinfo {author} {\bibfnamefont {Y.}~\bibnamefont
  {{Tsapras}}}, \bibinfo {author} {\bibfnamefont {D.}~\bibnamefont {{Valls-Gabaud}}},\ and\ \bibinfo {author} {\bibnamefont {{Point-Agape Collaboration}}},\ }\bibfield  {title} {\bibinfo {title} {{The Anomaly in the Candidate Microlensing Event PA-99-N2}},\ }\href {https://doi.org/10.1086/380820} {\bibfield  {journal} {\bibinfo  {journal} {APJ}\ }\textbf {\bibinfo {volume} {601}},\ \bibinfo {pages} {845} (\bibinfo {year} {2004})},\ \Eprint {https://arxiv.org/abs/astro-ph/0310457} {arXiv:astro-ph/0310457 [astro-ph]} \BibitemShut {NoStop}%
\bibitem [{\citenamefont {{Tuntsov}}\ \emph {et~al.}(2024)\citenamefont {{Tuntsov}}, \citenamefont {{Lewis}},\ and\ \citenamefont {{Walker}}}]{Tuntsov2024}%
  \BibitemOpen
  \bibfield  {author} {\bibinfo {author} {\bibfnamefont {A.~V.}\ \bibnamefont {{Tuntsov}}}, \bibinfo {author} {\bibfnamefont {G.~F.}\ \bibnamefont {{Lewis}}},\ and\ \bibinfo {author} {\bibfnamefont {M.~A.}\ \bibnamefont {{Walker}}},\ }\bibfield  {title} {\bibinfo {title} {{Free-floating 'planets' in the macrolensed quasar Q2237+0305}},\ }\href {https://doi.org/10.1093/mnras/stae133} {\bibfield  {journal} {\bibinfo  {journal} {MNRAS}\ }\textbf {\bibinfo {volume} {528}},\ \bibinfo {pages} {1979} (\bibinfo {year} {2024})},\ \Eprint {https://arxiv.org/abs/2401.05590} {arXiv:2401.05590 [astro-ph.CO]} \BibitemShut {NoStop}%
\bibitem [{\citenamefont {{Connor}}\ and\ \citenamefont {{Ravi}}(2023)}]{Connor2023}%
  \BibitemOpen
  \bibfield  {author} {\bibinfo {author} {\bibfnamefont {L.}~\bibnamefont {{Connor}}}\ and\ \bibinfo {author} {\bibfnamefont {V.}~\bibnamefont {{Ravi}}},\ }\bibfield  {title} {\bibinfo {title} {{Stellar prospects for FRB gravitational lensing}},\ }\href {https://doi.org/10.1093/mnras/stad667} {\bibfield  {journal} {\bibinfo  {journal} {MNRAS}\ }\textbf {\bibinfo {volume} {521}},\ \bibinfo {pages} {4024} (\bibinfo {year} {2023})},\ \Eprint {https://arxiv.org/abs/2206.14310} {arXiv:2206.14310 [astro-ph.CO]} \BibitemShut {NoStop}%
\bibitem [{\citenamefont {{Cassan}}\ \emph {et~al.}(2012)\citenamefont {{Cassan}}, \citenamefont {{Kubas}}, \citenamefont {{Beaulieu}}, \citenamefont {{Dominik}}, \citenamefont {{Horne}}, \citenamefont {{Greenhill}}, \citenamefont {{Wambsganss}}, \citenamefont {{Menzies}}, \citenamefont {{Williams}}, \citenamefont {{J{\o}rgensen}}, \citenamefont {{Udalski}}, \citenamefont {{Bennett}}, \citenamefont {{Albrow}}, \citenamefont {{Batista}}, \citenamefont {{Brillant}}, \citenamefont {{Caldwell}}, \citenamefont {{Cole}}, \citenamefont {{Coutures}}, \citenamefont {{Cook}}, \citenamefont {{Dieters}}, \citenamefont {{Dominis Prester}}, \citenamefont {{Donatowicz}}, \citenamefont {{Fouqu{\'e}}}, \citenamefont {{Hill}}, \citenamefont {{Kains}}, \citenamefont {{Kane}}, \citenamefont {{Marquette}}, \citenamefont {{Martin}}, \citenamefont {{Pollard}}, \citenamefont {{Sahu}}, \citenamefont {{Vinter}}, \citenamefont {{Warren}}, \citenamefont {{Watson}}, \citenamefont {{Zub}}, \citenamefont {{Sumi}}, \citenamefont
  {{Szyma{\'n}ski}}, \citenamefont {{Kubiak}}, \citenamefont {{Poleski}}, \citenamefont {{Soszynski}}, \citenamefont {{Ulaczyk}}, \citenamefont {{Pietrzy{\'n}ski}},\ and\ \citenamefont {{Wyrzykowski}}}]{2012Natur.481..167C}%
  \BibitemOpen
  \bibfield  {author} {\bibinfo {author} {\bibfnamefont {A.}~\bibnamefont {{Cassan}}}, \bibinfo {author} {\bibfnamefont {D.}~\bibnamefont {{Kubas}}}, \bibinfo {author} {\bibfnamefont {J.-P.}\ \bibnamefont {{Beaulieu}}}, \bibinfo {author} {\bibfnamefont {M.}~\bibnamefont {{Dominik}}}, \bibinfo {author} {\bibfnamefont {K.}~\bibnamefont {{Horne}}}, \bibinfo {author} {\bibfnamefont {J.}~\bibnamefont {{Greenhill}}}, \bibinfo {author} {\bibfnamefont {J.}~\bibnamefont {{Wambsganss}}}, \bibinfo {author} {\bibfnamefont {J.}~\bibnamefont {{Menzies}}}, \bibinfo {author} {\bibfnamefont {A.}~\bibnamefont {{Williams}}}, \bibinfo {author} {\bibfnamefont {U.~G.}\ \bibnamefont {{J{\o}rgensen}}}, \bibinfo {author} {\bibfnamefont {A.}~\bibnamefont {{Udalski}}}, \bibinfo {author} {\bibfnamefont {D.~P.}\ \bibnamefont {{Bennett}}}, \bibinfo {author} {\bibfnamefont {M.~D.}\ \bibnamefont {{Albrow}}}, \bibinfo {author} {\bibfnamefont {V.}~\bibnamefont {{Batista}}}, \bibinfo {author} {\bibfnamefont {S.}~\bibnamefont {{Brillant}}},
  \bibinfo {author} {\bibfnamefont {J.~A.~R.}\ \bibnamefont {{Caldwell}}}, \bibinfo {author} {\bibfnamefont {A.}~\bibnamefont {{Cole}}}, \bibinfo {author} {\bibfnamefont {C.}~\bibnamefont {{Coutures}}}, \bibinfo {author} {\bibfnamefont {K.~H.}\ \bibnamefont {{Cook}}}, \bibinfo {author} {\bibfnamefont {S.}~\bibnamefont {{Dieters}}}, \bibinfo {author} {\bibfnamefont {D.}~\bibnamefont {{Dominis Prester}}}, \bibinfo {author} {\bibfnamefont {J.}~\bibnamefont {{Donatowicz}}}, \bibinfo {author} {\bibfnamefont {P.}~\bibnamefont {{Fouqu{\'e}}}}, \bibinfo {author} {\bibfnamefont {K.}~\bibnamefont {{Hill}}}, \bibinfo {author} {\bibfnamefont {N.}~\bibnamefont {{Kains}}}, \bibinfo {author} {\bibfnamefont {S.}~\bibnamefont {{Kane}}}, \bibinfo {author} {\bibfnamefont {J.-B.}\ \bibnamefont {{Marquette}}}, \bibinfo {author} {\bibfnamefont {R.}~\bibnamefont {{Martin}}}, \bibinfo {author} {\bibfnamefont {K.~R.}\ \bibnamefont {{Pollard}}}, \bibinfo {author} {\bibfnamefont {K.~C.}\ \bibnamefont {{Sahu}}}, \bibinfo {author}
  {\bibfnamefont {C.}~\bibnamefont {{Vinter}}}, \bibinfo {author} {\bibfnamefont {D.}~\bibnamefont {{Warren}}}, \bibinfo {author} {\bibfnamefont {B.}~\bibnamefont {{Watson}}}, \bibinfo {author} {\bibfnamefont {M.}~\bibnamefont {{Zub}}}, \bibinfo {author} {\bibfnamefont {T.}~\bibnamefont {{Sumi}}}, \bibinfo {author} {\bibfnamefont {M.~K.}\ \bibnamefont {{Szyma{\'n}ski}}}, \bibinfo {author} {\bibfnamefont {M.}~\bibnamefont {{Kubiak}}}, \bibinfo {author} {\bibfnamefont {R.}~\bibnamefont {{Poleski}}}, \bibinfo {author} {\bibfnamefont {I.}~\bibnamefont {{Soszynski}}}, \bibinfo {author} {\bibfnamefont {K.}~\bibnamefont {{Ulaczyk}}}, \bibinfo {author} {\bibfnamefont {G.}~\bibnamefont {{Pietrzy{\'n}ski}}},\ and\ \bibinfo {author} {\bibfnamefont {{\L}.}~\bibnamefont {{Wyrzykowski}}},\ }\bibfield  {title} {\bibinfo {title} {{One or more bound planets per Milky Way star from microlensing observations}},\ }\href {https://doi.org/10.1038/nature10684} {\bibfield  {journal} {\bibinfo  {journal} {NAT}\ }\textbf {\bibinfo
  {volume} {481}},\ \bibinfo {pages} {167} (\bibinfo {year} {2012})},\ \Eprint {https://arxiv.org/abs/1202.0903} {arXiv:1202.0903 [astro-ph.EP]} \BibitemShut {NoStop}%
\bibitem [{\citenamefont {{Leung}}\ \emph {et~al.}(2022)\citenamefont {{Leung}}, \citenamefont {{Kader}}, \citenamefont {{Masui}}, \citenamefont {{Dobbs}}, \citenamefont {{Michilli}}, \citenamefont {{Mena-Parra}}, \citenamefont {{Mckinven}}, \citenamefont {{Ng}}, \citenamefont {{Bandura}}, \citenamefont {{Bhardwaj}}, \citenamefont {{Brar}}, \citenamefont {{Cassanelli}}, \citenamefont {{Chawla}}, \citenamefont {{Dong}}, \citenamefont {{Good}}, \citenamefont {{Kaspi}}, \citenamefont {{Lanman}}, \citenamefont {{Lin}}, \citenamefont {{Meyers}}, \citenamefont {{Pearlman}}, \citenamefont {{Pen}}, \citenamefont {{Petroff}}, \citenamefont {{Pleunis}}, \citenamefont {{Rafiei-Ravandi}}, \citenamefont {{Rahman}}, \citenamefont {{Sanghavi}}, \citenamefont {{Scholz}}, \citenamefont {{Shin}}, \citenamefont {{Siegel}}, \citenamefont {{Smith}}, \citenamefont {{Stairs}}, \citenamefont {{Tendulkar}},\ and\ \citenamefont {{Vanderlinde}}}]{Leung2022}%
  \BibitemOpen
  \bibfield  {author} {\bibinfo {author} {\bibfnamefont {C.}~\bibnamefont {{Leung}}}, \bibinfo {author} {\bibfnamefont {Z.}~\bibnamefont {{Kader}}}, \bibinfo {author} {\bibfnamefont {K.~W.}\ \bibnamefont {{Masui}}}, \bibinfo {author} {\bibfnamefont {M.}~\bibnamefont {{Dobbs}}}, \bibinfo {author} {\bibfnamefont {D.}~\bibnamefont {{Michilli}}}, \bibinfo {author} {\bibfnamefont {J.}~\bibnamefont {{Mena-Parra}}}, \bibinfo {author} {\bibfnamefont {R.}~\bibnamefont {{Mckinven}}}, \bibinfo {author} {\bibfnamefont {C.}~\bibnamefont {{Ng}}}, \bibinfo {author} {\bibfnamefont {K.}~\bibnamefont {{Bandura}}}, \bibinfo {author} {\bibfnamefont {M.}~\bibnamefont {{Bhardwaj}}}, \bibinfo {author} {\bibfnamefont {C.}~\bibnamefont {{Brar}}}, \bibinfo {author} {\bibfnamefont {T.}~\bibnamefont {{Cassanelli}}}, \bibinfo {author} {\bibfnamefont {P.}~\bibnamefont {{Chawla}}}, \bibinfo {author} {\bibfnamefont {F.~A.}\ \bibnamefont {{Dong}}}, \bibinfo {author} {\bibfnamefont {D.}~\bibnamefont {{Good}}}, \bibinfo {author} {\bibfnamefont
  {V.}~\bibnamefont {{Kaspi}}}, \bibinfo {author} {\bibfnamefont {A.~E.}\ \bibnamefont {{Lanman}}}, \bibinfo {author} {\bibfnamefont {H.-H.}\ \bibnamefont {{Lin}}}, \bibinfo {author} {\bibfnamefont {B.~W.}\ \bibnamefont {{Meyers}}}, \bibinfo {author} {\bibfnamefont {A.~B.}\ \bibnamefont {{Pearlman}}}, \bibinfo {author} {\bibfnamefont {U.-L.}\ \bibnamefont {{Pen}}}, \bibinfo {author} {\bibfnamefont {E.}~\bibnamefont {{Petroff}}}, \bibinfo {author} {\bibfnamefont {Z.}~\bibnamefont {{Pleunis}}}, \bibinfo {author} {\bibfnamefont {M.}~\bibnamefont {{Rafiei-Ravandi}}}, \bibinfo {author} {\bibfnamefont {M.}~\bibnamefont {{Rahman}}}, \bibinfo {author} {\bibfnamefont {P.}~\bibnamefont {{Sanghavi}}}, \bibinfo {author} {\bibfnamefont {P.}~\bibnamefont {{Scholz}}}, \bibinfo {author} {\bibfnamefont {K.}~\bibnamefont {{Shin}}}, \bibinfo {author} {\bibfnamefont {S.}~\bibnamefont {{Siegel}}}, \bibinfo {author} {\bibfnamefont {K.~M.}\ \bibnamefont {{Smith}}}, \bibinfo {author} {\bibfnamefont {I.}~\bibnamefont {{Stairs}}},
  \bibinfo {author} {\bibfnamefont {S.~P.}\ \bibnamefont {{Tendulkar}}},\ and\ \bibinfo {author} {\bibfnamefont {K.}~\bibnamefont {{Vanderlinde}}},\ }\bibfield  {title} {\bibinfo {title} {{Constraining primordial black holes using fast radio burst gravitational-lens interferometry with CHIME/FRB}},\ }\href {https://doi.org/10.1103/PhysRevD.106.043017} {\bibfield  {journal} {\bibinfo  {journal} {PRD}\ }\textbf {\bibinfo {volume} {106}},\ \bibinfo {eid} {043017} (\bibinfo {year} {2022})},\ \Eprint {https://arxiv.org/abs/2204.06001} {arXiv:2204.06001 [astro-ph.HE]} \BibitemShut {NoStop}%
\bibitem [{\citenamefont {{Mordasini}}\ \emph {et~al.}(2009)\citenamefont {{Mordasini}}, \citenamefont {{Alibert}},\ and\ \citenamefont {{Benz}}}]{2009A&A...501.1139M}%
  \BibitemOpen
  \bibfield  {author} {\bibinfo {author} {\bibfnamefont {C.}~\bibnamefont {{Mordasini}}}, \bibinfo {author} {\bibfnamefont {Y.}~\bibnamefont {{Alibert}}},\ and\ \bibinfo {author} {\bibfnamefont {W.}~\bibnamefont {{Benz}}},\ }\bibfield  {title} {\bibinfo {title} {{Extrasolar planet population synthesis. I. Method, formation tracks, and mass-distance distribution}},\ }\href {https://doi.org/10.1051/0004-6361/200810301} {\bibfield  {journal} {\bibinfo  {journal} {AAP}\ }\textbf {\bibinfo {volume} {501}},\ \bibinfo {pages} {1139} (\bibinfo {year} {2009})},\ \Eprint {https://arxiv.org/abs/0904.2524} {arXiv:0904.2524 [astro-ph.EP]} \BibitemShut {NoStop}%
\bibitem [{\citenamefont {{El-Badry}}(2024)}]{Kareem2024NewAR..9801694E}%
  \BibitemOpen
  \bibfield  {author} {\bibinfo {author} {\bibfnamefont {K.}~\bibnamefont {{El-Badry}}},\ }\bibfield  {title} {\bibinfo {title} {{Gaia's binary star renaissance}},\ }\href {https://doi.org/10.1016/j.newar.2024.101694} {\bibfield  {journal} {\bibinfo  {journal} {NAR}\ }\textbf {\bibinfo {volume} {98}},\ \bibinfo {eid} {101694} (\bibinfo {year} {2024})},\ \Eprint {https://arxiv.org/abs/2403.12146} {arXiv:2403.12146 [astro-ph.SR]} \BibitemShut {NoStop}%
\bibitem [{\citenamefont {{Mr{\'o}z}}\ \emph {et~al.}(2019)\citenamefont {{Mr{\'o}z}}, \citenamefont {{Udalski}}, \citenamefont {{Skowron}}, \citenamefont {{Szyma{\'n}ski}}, \citenamefont {{Soszy{\'n}ski}}, \citenamefont {{Wyrzykowski}}, \citenamefont {{Pietrukowicz}}, \citenamefont {{Koz{\l}owski}}, \citenamefont {{Poleski}}, \citenamefont {{Ulaczyk}}, \citenamefont {{Rybicki}},\ and\ \citenamefont {{Iwanek}}}]{Mroz2019ApJS..244...29M}%
  \BibitemOpen
  \bibfield  {author} {\bibinfo {author} {\bibfnamefont {P.}~\bibnamefont {{Mr{\'o}z}}}, \bibinfo {author} {\bibfnamefont {A.}~\bibnamefont {{Udalski}}}, \bibinfo {author} {\bibfnamefont {J.}~\bibnamefont {{Skowron}}}, \bibinfo {author} {\bibfnamefont {M.~K.}\ \bibnamefont {{Szyma{\'n}ski}}}, \bibinfo {author} {\bibfnamefont {I.}~\bibnamefont {{Soszy{\'n}ski}}}, \bibinfo {author} {\bibfnamefont {{\L}.}~\bibnamefont {{Wyrzykowski}}}, \bibinfo {author} {\bibfnamefont {P.}~\bibnamefont {{Pietrukowicz}}}, \bibinfo {author} {\bibfnamefont {S.}~\bibnamefont {{Koz{\l}owski}}}, \bibinfo {author} {\bibfnamefont {R.}~\bibnamefont {{Poleski}}}, \bibinfo {author} {\bibfnamefont {K.}~\bibnamefont {{Ulaczyk}}}, \bibinfo {author} {\bibfnamefont {K.}~\bibnamefont {{Rybicki}}},\ and\ \bibinfo {author} {\bibfnamefont {P.}~\bibnamefont {{Iwanek}}},\ }\bibfield  {title} {\bibinfo {title} {{Microlensing Optical Depth and Event Rate toward the Galactic Bulge from 8 yr of OGLE-IV Observations}},\ }\href
  {https://doi.org/10.3847/1538-4365/ab426b} {\bibfield  {journal} {\bibinfo  {journal} {APJS}\ }\textbf {\bibinfo {volume} {244}},\ \bibinfo {eid} {29} (\bibinfo {year} {2019})},\ \Eprint {https://arxiv.org/abs/1906.02210} {arXiv:1906.02210 [astro-ph.SR]} \BibitemShut {NoStop}%
\bibitem [{\citenamefont {{Jow}}\ \emph {et~al.}(2020)\citenamefont {{Jow}}, \citenamefont {{Foreman}}, \citenamefont {{Pen}},\ and\ \citenamefont {{Zhu}}}]{Jow2020MNRAS.497.4956J}%
  \BibitemOpen
  \bibfield  {author} {\bibinfo {author} {\bibfnamefont {D.~L.}\ \bibnamefont {{Jow}}}, \bibinfo {author} {\bibfnamefont {S.}~\bibnamefont {{Foreman}}}, \bibinfo {author} {\bibfnamefont {U.-L.}\ \bibnamefont {{Pen}}},\ and\ \bibinfo {author} {\bibfnamefont {W.}~\bibnamefont {{Zhu}}},\ }\bibfield  {title} {\bibinfo {title} {{Wave effects in the microlensing of pulsars and FRBs by point masses}},\ }\href {https://doi.org/10.1093/mnras/staa2230} {\bibfield  {journal} {\bibinfo  {journal} {MNRAS}\ }\textbf {\bibinfo {volume} {497}},\ \bibinfo {pages} {4956} (\bibinfo {year} {2020})},\ \Eprint {https://arxiv.org/abs/2002.01570} {arXiv:2002.01570 [astro-ph.HE]} \BibitemShut {NoStop}%
\bibitem [{\citenamefont {{Hallinan}}\ \emph {et~al.}(2019)\citenamefont {{Hallinan}}, \citenamefont {{Ravi}}, \citenamefont {{Weinreb}}, \citenamefont {{Kocz}}, \citenamefont {{Huang}}, \citenamefont {{Woody}}, \citenamefont {{Lamb}}, \citenamefont {{D'Addario}}, \citenamefont {{Catha}}, \citenamefont {{Law}}, \citenamefont {{Kulkarni}}, \citenamefont {{Phinney}}, \citenamefont {{Eastwood}}, \citenamefont {{Bouman}}, \citenamefont {{McLaughlin}}, \citenamefont {{Ransom}}, \citenamefont {{Siemens}}, \citenamefont {{Cordes}}, \citenamefont {{Lynch}}, \citenamefont {{Kaplan}}, \citenamefont {{Brazier}}, \citenamefont {{Bhatnagar}}, \citenamefont {{Myers}}, \citenamefont {{Walter}},\ and\ \citenamefont {{Gaensler}}}]{DSA2019BAAS...51g.255H}%
  \BibitemOpen
  \bibfield  {author} {\bibinfo {author} {\bibfnamefont {G.}~\bibnamefont {{Hallinan}}}, \bibinfo {author} {\bibfnamefont {V.}~\bibnamefont {{Ravi}}}, \bibinfo {author} {\bibfnamefont {S.}~\bibnamefont {{Weinreb}}}, \bibinfo {author} {\bibfnamefont {J.}~\bibnamefont {{Kocz}}}, \bibinfo {author} {\bibfnamefont {Y.}~\bibnamefont {{Huang}}}, \bibinfo {author} {\bibfnamefont {D.~P.}\ \bibnamefont {{Woody}}}, \bibinfo {author} {\bibfnamefont {J.}~\bibnamefont {{Lamb}}}, \bibinfo {author} {\bibfnamefont {L.}~\bibnamefont {{D'Addario}}}, \bibinfo {author} {\bibfnamefont {M.}~\bibnamefont {{Catha}}}, \bibinfo {author} {\bibfnamefont {C.}~\bibnamefont {{Law}}}, \bibinfo {author} {\bibfnamefont {S.~R.}\ \bibnamefont {{Kulkarni}}}, \bibinfo {author} {\bibfnamefont {E.~S.}\ \bibnamefont {{Phinney}}}, \bibinfo {author} {\bibfnamefont {M.~W.}\ \bibnamefont {{Eastwood}}}, \bibinfo {author} {\bibfnamefont {K.}~\bibnamefont {{Bouman}}}, \bibinfo {author} {\bibfnamefont {M.}~\bibnamefont {{McLaughlin}}}, \bibinfo {author}
  {\bibfnamefont {S.}~\bibnamefont {{Ransom}}}, \bibinfo {author} {\bibfnamefont {X.}~\bibnamefont {{Siemens}}}, \bibinfo {author} {\bibfnamefont {J.}~\bibnamefont {{Cordes}}}, \bibinfo {author} {\bibfnamefont {R.}~\bibnamefont {{Lynch}}}, \bibinfo {author} {\bibfnamefont {D.}~\bibnamefont {{Kaplan}}}, \bibinfo {author} {\bibfnamefont {A.}~\bibnamefont {{Brazier}}}, \bibinfo {author} {\bibfnamefont {S.}~\bibnamefont {{Bhatnagar}}}, \bibinfo {author} {\bibfnamefont {S.}~\bibnamefont {{Myers}}}, \bibinfo {author} {\bibfnamefont {F.}~\bibnamefont {{Walter}}},\ and\ \bibinfo {author} {\bibfnamefont {B.}~\bibnamefont {{Gaensler}}},\ }\bibfield  {title} {\bibinfo {title} {{The DSA-2000 {\textemdash} A Radio Survey Camera}},\ }in\ \href {https://doi.org/10.48550/arXiv.1907.07648} {\emph {\bibinfo {booktitle} {Bulletin of the American Astronomical Society}}},\ Vol.~\bibinfo {volume} {51}\ (\bibinfo {year} {2019})\ p.\ \bibinfo {pages} {255},\ \Eprint {https://arxiv.org/abs/1907.07648} {arXiv:1907.07648 [astro-ph.IM]}
  \BibitemShut {NoStop}%
\bibitem [{\citenamefont {{Vanderlinde}}\ \emph {et~al.}(2019)\citenamefont {{Vanderlinde}}, \citenamefont {{Liu}}, \citenamefont {{Gaensler}}, \citenamefont {{Bond}}, \citenamefont {{Hinshaw}}, \citenamefont {{Ng}}, \citenamefont {{Chiang}}, \citenamefont {{Stairs}}, \citenamefont {{Brown}}, \citenamefont {{Sievers}}, \citenamefont {{Mena}}, \citenamefont {{Smith}}, \citenamefont {{Bandura}}, \citenamefont {{Masui}}, \citenamefont {{Spekkens}}, \citenamefont {{Belostotski}}, \citenamefont {{Dobbs}}, \citenamefont {{Turok}}, \citenamefont {{Boyle}}, \citenamefont {{Rupen}}, \citenamefont {{Landecker}}, \citenamefont {{Pen}},\ and\ \citenamefont {{Kaspi}}}]{CHORD2019clrp.2020...28V}%
  \BibitemOpen
  \bibfield  {author} {\bibinfo {author} {\bibfnamefont {K.}~\bibnamefont {{Vanderlinde}}}, \bibinfo {author} {\bibfnamefont {A.}~\bibnamefont {{Liu}}}, \bibinfo {author} {\bibfnamefont {B.}~\bibnamefont {{Gaensler}}}, \bibinfo {author} {\bibfnamefont {D.}~\bibnamefont {{Bond}}}, \bibinfo {author} {\bibfnamefont {G.}~\bibnamefont {{Hinshaw}}}, \bibinfo {author} {\bibfnamefont {C.}~\bibnamefont {{Ng}}}, \bibinfo {author} {\bibfnamefont {C.}~\bibnamefont {{Chiang}}}, \bibinfo {author} {\bibfnamefont {I.}~\bibnamefont {{Stairs}}}, \bibinfo {author} {\bibfnamefont {J.-A.}\ \bibnamefont {{Brown}}}, \bibinfo {author} {\bibfnamefont {J.}~\bibnamefont {{Sievers}}}, \bibinfo {author} {\bibfnamefont {J.}~\bibnamefont {{Mena}}}, \bibinfo {author} {\bibfnamefont {K.}~\bibnamefont {{Smith}}}, \bibinfo {author} {\bibfnamefont {K.}~\bibnamefont {{Bandura}}}, \bibinfo {author} {\bibfnamefont {K.}~\bibnamefont {{Masui}}}, \bibinfo {author} {\bibfnamefont {K.}~\bibnamefont {{Spekkens}}}, \bibinfo {author} {\bibfnamefont
  {L.}~\bibnamefont {{Belostotski}}}, \bibinfo {author} {\bibfnamefont {M.}~\bibnamefont {{Dobbs}}}, \bibinfo {author} {\bibfnamefont {N.}~\bibnamefont {{Turok}}}, \bibinfo {author} {\bibfnamefont {P.}~\bibnamefont {{Boyle}}}, \bibinfo {author} {\bibfnamefont {M.}~\bibnamefont {{Rupen}}}, \bibinfo {author} {\bibfnamefont {T.}~\bibnamefont {{Landecker}}}, \bibinfo {author} {\bibfnamefont {U.-L.}\ \bibnamefont {{Pen}}},\ and\ \bibinfo {author} {\bibfnamefont {V.}~\bibnamefont {{Kaspi}}},\ }\bibfield  {title} {\bibinfo {title} {{The Canadian Hydrogen Observatory and Radio-transient Detector (CHORD)}},\ }in\ \href {https://doi.org/10.5281/zenodo.3765414} {\emph {\bibinfo {booktitle} {Canadian Long Range Plan for Astronomy and Astrophysics White Papers}}},\ Vol.\ \bibinfo {volume} {2020}\ (\bibinfo {year} {2019})\ p.~\bibinfo {pages} {28},\ \Eprint {https://arxiv.org/abs/1911.01777} {arXiv:1911.01777 [astro-ph.IM]} \BibitemShut {NoStop}%
\bibitem [{\citenamefont {{Lin}}\ \emph {et~al.}(2022)\citenamefont {{Lin}}, \citenamefont {{Lin}}, \citenamefont {{Li}}, \citenamefont {{Tseng}}, \citenamefont {{Jiang}}, \citenamefont {{Wang}}, \citenamefont {{Cheng}}, \citenamefont {{Pen}}, \citenamefont {{Chen}}, \citenamefont {{Chen}}, \citenamefont {{Chen}}, \citenamefont {{Goto}}, \citenamefont {{Hashimoto}}, \citenamefont {{Hwang}}, \citenamefont {{King}}, \citenamefont {{Kubo}}, \citenamefont {{Kuo}}, \citenamefont {{Mills}}, \citenamefont {{Nam}}, \citenamefont {{Oshiro}}, \citenamefont {{Shen}}, \citenamefont {{Tseng}}, \citenamefont {{Wang}}, \citenamefont {{Wu}}, \citenamefont {{Bower}}, \citenamefont {{Chang}}, \citenamefont {{Chen}}, \citenamefont {{Chen}}, \citenamefont {{Chiang}}, \citenamefont {{Fedynitch}}, \citenamefont {{Gusinskaia}}, \citenamefont {{Ho}}, \citenamefont {{Hsiao}}, \citenamefont {{Hu}}, \citenamefont {{Huang}}, \citenamefont {{J{\'a}uregui Garc{\'\i}a}}, \citenamefont {{Kim}}, \citenamefont {{Kuo}}, \citenamefont {{Ling}},
  \citenamefont {{On}}, \citenamefont {{Peterson}}, \citenamefont {{R. Raquel}}, \citenamefont {{Su}}, \citenamefont {{Uno}}, \citenamefont {{Wu}}, \citenamefont {{Yamasaki}},\ and\ \citenamefont {{Zhu}}}]{BURSTT2022PASP..134i4106L}%
  \BibitemOpen
  \bibfield  {author} {\bibinfo {author} {\bibfnamefont {H.-H.}\ \bibnamefont {{Lin}}}, \bibinfo {author} {\bibfnamefont {K.-y.}\ \bibnamefont {{Lin}}}, \bibinfo {author} {\bibfnamefont {C.-T.}\ \bibnamefont {{Li}}}, \bibinfo {author} {\bibfnamefont {Y.-H.}\ \bibnamefont {{Tseng}}}, \bibinfo {author} {\bibfnamefont {H.}~\bibnamefont {{Jiang}}}, \bibinfo {author} {\bibfnamefont {J.-H.}\ \bibnamefont {{Wang}}}, \bibinfo {author} {\bibfnamefont {J.-C.}\ \bibnamefont {{Cheng}}}, \bibinfo {author} {\bibfnamefont {U.-L.}\ \bibnamefont {{Pen}}}, \bibinfo {author} {\bibfnamefont {M.-T.}\ \bibnamefont {{Chen}}}, \bibinfo {author} {\bibfnamefont {P.}~\bibnamefont {{Chen}}}, \bibinfo {author} {\bibfnamefont {Y.}~\bibnamefont {{Chen}}}, \bibinfo {author} {\bibfnamefont {T.}~\bibnamefont {{Goto}}}, \bibinfo {author} {\bibfnamefont {T.}~\bibnamefont {{Hashimoto}}}, \bibinfo {author} {\bibfnamefont {Y.-J.}\ \bibnamefont {{Hwang}}}, \bibinfo {author} {\bibfnamefont {S.-K.}\ \bibnamefont {{King}}}, \bibinfo {author}
  {\bibfnamefont {D.}~\bibnamefont {{Kubo}}}, \bibinfo {author} {\bibfnamefont {C.-Y.}\ \bibnamefont {{Kuo}}}, \bibinfo {author} {\bibfnamefont {A.}~\bibnamefont {{Mills}}}, \bibinfo {author} {\bibfnamefont {J.}~\bibnamefont {{Nam}}}, \bibinfo {author} {\bibfnamefont {P.}~\bibnamefont {{Oshiro}}}, \bibinfo {author} {\bibfnamefont {C.-S.}\ \bibnamefont {{Shen}}}, \bibinfo {author} {\bibfnamefont {H.-C.}\ \bibnamefont {{Tseng}}}, \bibinfo {author} {\bibfnamefont {S.-H.}\ \bibnamefont {{Wang}}}, \bibinfo {author} {\bibfnamefont {V.~F.-S.}\ \bibnamefont {{Wu}}}, \bibinfo {author} {\bibfnamefont {G.}~\bibnamefont {{Bower}}}, \bibinfo {author} {\bibfnamefont {S.-H.}\ \bibnamefont {{Chang}}}, \bibinfo {author} {\bibfnamefont {P.-A.}\ \bibnamefont {{Chen}}}, \bibinfo {author} {\bibfnamefont {Y.-C.}\ \bibnamefont {{Chen}}}, \bibinfo {author} {\bibfnamefont {Y.-K.}\ \bibnamefont {{Chiang}}}, \bibinfo {author} {\bibfnamefont {A.}~\bibnamefont {{Fedynitch}}}, \bibinfo {author} {\bibfnamefont {N.}~\bibnamefont
  {{Gusinskaia}}}, \bibinfo {author} {\bibfnamefont {S.~C.-C.}\ \bibnamefont {{Ho}}}, \bibinfo {author} {\bibfnamefont {T.~Y.-Y.}\ \bibnamefont {{Hsiao}}}, \bibinfo {author} {\bibfnamefont {C.-P.}\ \bibnamefont {{Hu}}}, \bibinfo {author} {\bibfnamefont {Y.~D.}\ \bibnamefont {{Huang}}}, \bibinfo {author} {\bibfnamefont {J.~M.}\ \bibnamefont {{J{\'a}uregui Garc{\'\i}a}}}, \bibinfo {author} {\bibfnamefont {S.~J.}\ \bibnamefont {{Kim}}}, \bibinfo {author} {\bibfnamefont {C.-Y.}\ \bibnamefont {{Kuo}}}, \bibinfo {author} {\bibfnamefont {D.~F.-J.}\ \bibnamefont {{Ling}}}, \bibinfo {author} {\bibfnamefont {A.~Y.~L.}\ \bibnamefont {{On}}}, \bibinfo {author} {\bibfnamefont {J.~B.}\ \bibnamefont {{Peterson}}}, \bibinfo {author} {\bibfnamefont {B.~J.}\ \bibnamefont {{R. Raquel}}}, \bibinfo {author} {\bibfnamefont {S.-C.}\ \bibnamefont {{Su}}}, \bibinfo {author} {\bibfnamefont {Y.}~\bibnamefont {{Uno}}}, \bibinfo {author} {\bibfnamefont {C.~K.-W.}\ \bibnamefont {{Wu}}}, \bibinfo {author} {\bibfnamefont {S.}~\bibnamefont
  {{Yamasaki}}},\ and\ \bibinfo {author} {\bibfnamefont {H.-M.}\ \bibnamefont {{Zhu}}},\ }\bibfield  {title} {\bibinfo {title} {{BURSTT: Bustling Universe Radio Survey Telescope in Taiwan}},\ }\href {https://doi.org/10.1088/1538-3873/ac8f71} {\bibfield  {journal} {\bibinfo  {journal} {PASP}\ }\textbf {\bibinfo {volume} {134}},\ \bibinfo {eid} {094106} (\bibinfo {year} {2022})},\ \Eprint {https://arxiv.org/abs/2206.08983} {arXiv:2206.08983 [astro-ph.IM]} \BibitemShut {NoStop}%
\bibitem [{\citenamefont {{Luo}}\ \emph {et~al.}(2024)\citenamefont {{Luo}}, \citenamefont {{Ekers}}, \citenamefont {{Hobbs}}, \citenamefont {{Dunning}}, \citenamefont {{James}}, \citenamefont {{Lower}}, \citenamefont {{Gupta}}, \citenamefont {{Zic}}, \citenamefont {{Sokolowski}}, \citenamefont {{Phillips}}, \citenamefont {{Deller}},\ and\ \citenamefont {{Staveley-Smith}}}]{Luo2024PASA...41..109L}%
  \BibitemOpen
  \bibfield  {author} {\bibinfo {author} {\bibfnamefont {R.}~\bibnamefont {{Luo}}}, \bibinfo {author} {\bibfnamefont {R.}~\bibnamefont {{Ekers}}}, \bibinfo {author} {\bibfnamefont {G.}~\bibnamefont {{Hobbs}}}, \bibinfo {author} {\bibfnamefont {A.}~\bibnamefont {{Dunning}}}, \bibinfo {author} {\bibfnamefont {C.}~\bibnamefont {{James}}}, \bibinfo {author} {\bibfnamefont {M.}~\bibnamefont {{Lower}}}, \bibinfo {author} {\bibfnamefont {V.}~\bibnamefont {{Gupta}}}, \bibinfo {author} {\bibfnamefont {A.}~\bibnamefont {{Zic}}}, \bibinfo {author} {\bibfnamefont {M.}~\bibnamefont {{Sokolowski}}}, \bibinfo {author} {\bibfnamefont {C.}~\bibnamefont {{Phillips}}}, \bibinfo {author} {\bibfnamefont {A.}~\bibnamefont {{Deller}}},\ and\ \bibinfo {author} {\bibfnamefont {L.}~\bibnamefont {{Staveley-Smith}}},\ }\bibfield  {title} {\bibinfo {title} {{A fast radio burst monitor with a compact all-sky phased array (CASPA)}},\ }\href {https://doi.org/10.1017/pasa.2024.108} {\bibfield  {journal} {\bibinfo  {journal} {PASA}\ }\textbf
  {\bibinfo {volume} {41}},\ \bibinfo {eid} {e109} (\bibinfo {year} {2024})},\ \Eprint {https://arxiv.org/abs/2405.07439} {arXiv:2405.07439 [astro-ph.IM]} \BibitemShut {NoStop}%
\bibitem [{\citenamefont {{Hashimoto}}\ \emph {et~al.}(2020)\citenamefont {{Hashimoto}}, \citenamefont {{Goto}}, \citenamefont {{On}}, \citenamefont {{Lu}}, \citenamefont {{Santos}}, \citenamefont {{Ho}}, \citenamefont {{Wang}}, \citenamefont {{Kim}},\ and\ \citenamefont {{Hsiao}}}]{SKArate2020MNRAS.497.4107H}%
  \BibitemOpen
  \bibfield  {author} {\bibinfo {author} {\bibfnamefont {T.}~\bibnamefont {{Hashimoto}}}, \bibinfo {author} {\bibfnamefont {T.}~\bibnamefont {{Goto}}}, \bibinfo {author} {\bibfnamefont {A.~Y.~L.}\ \bibnamefont {{On}}}, \bibinfo {author} {\bibfnamefont {T.-Y.}\ \bibnamefont {{Lu}}}, \bibinfo {author} {\bibfnamefont {D.~J.~D.}\ \bibnamefont {{Santos}}}, \bibinfo {author} {\bibfnamefont {S.~C.-C.}\ \bibnamefont {{Ho}}}, \bibinfo {author} {\bibfnamefont {T.-W.}\ \bibnamefont {{Wang}}}, \bibinfo {author} {\bibfnamefont {S.~J.}\ \bibnamefont {{Kim}}},\ and\ \bibinfo {author} {\bibfnamefont {T.~Y.-Y.}\ \bibnamefont {{Hsiao}}},\ }\bibfield  {title} {\bibinfo {title} {{Fast radio bursts to be detected with the Square Kilometre Array}},\ }\href {https://doi.org/10.1093/mnras/staa2238} {\bibfield  {journal} {\bibinfo  {journal} {MNRAS}\ }\textbf {\bibinfo {volume} {497}},\ \bibinfo {pages} {4107} (\bibinfo {year} {2020})},\ \Eprint {https://arxiv.org/abs/2008.00007} {arXiv:2008.00007 [astro-ph.HE]} \BibitemShut
  {NoStop}%
\end{thebibliography}%

\end{document}